\documentclass[a4paper,12pt]{article}            
\usepackage{jheppub}
\pdfoutput=1

\usepackage[utf8]{inputenc}
\usepackage{amsmath,amssymb}
\usepackage{xcolor}
\usepackage{tensor}
\usepackage{graphicx}
\usepackage{subcaption}
\usepackage{xurl} 
\usepackage{hyperref}
\usepackage{bbm}
\usepackage{cancel}
\usepackage{mathrsfs}

\title{The price of locality: Maxwell-Cattaneo charge transport in Schwinger-Keldysh effective field theory}
\author[1,2]{Andrea Amoretti,}
\author[1,2]{Matteo Anselmi,} 
\author[1,2]{Daniel K. Brattan.}

\emailAdd{andrea.amoretti@ge.infn.it}
\emailAdd{matteo.anselmi@ge.infn.it}
\emailAdd{danny.brattan@gmail.com}

\affiliation[1]{Dipartimento di Fisica, Universit\`a di Genova,
via Dodecaneso 33, I-16146, Genova, Italy}
\affiliation[2]{I.N.F.N. - Sezione di Genova, via Dodecaneso 33, I-16146, Genova, Italy}

\begin{abstract}
{\ We determine the conditions under which the nonlinear Maxwell-Cattaneo theory of charge transport admits a local Gaussian Schwinger-Keldysh embedding with a particular modified dynamical KMS symmetry. We show that treating the additional vector as intrinsically dissipative fundamentally modifies the entropy-current analysis: the hydrostatic generating functional no longer fixes all entropy-current improvements, which must instead be supplied independently. Subsequently we discuss the additional constraints not contained in the hydrodynamic analysis that are imposed by the modified KMS transformation becoming a symmetry of the action. We both interpret and supply the corresponding stochastic quasi-hydrodynamics.}
\end{abstract}
\begin{document}

\maketitle

\section{Introduction }

{\noindent Hydrodynamics provides the universal effective description of systems close to local equilibrium. In recent years, the Schwinger-Keldysh formalism \cite{Haehl:2014zda,Haehl:2015uoc,Haehl:2016pec,Crossley:2015evo,Glorioso:2017fpd,Glorioso:2016gsa,Gao:2017bqf,Haehl:2018lcu,Glorioso:2018wxw,Jensen:2018hse} has provided a systematic framework for constructing the corresponding stochastic effective theories, making explicit the relationship between dissipation, thermal fluctuations and dynamical KMS symmetry. While this relationship is well understood for conventional hydrodynamics, much less is known when the effective theory contains additional non-hydrodynamic degrees of freedom responsible for higher poles in retarded Green's functions. A familiar example is Maxwell-Cattaneo theory (see \cite{Joseph:1989zz} for a review), which augments diffusive charge transport by a relaxation mode and thereby captures a leading non-hydrodynamic pole of the conductivity when present.}

{\ A common feature of many microscopic systems is the existence of more than the hydrodynamic mode in the retarded correlator \cite{Amoretti:2025kem}. These non-hydrodynamic pole data are themselves non-trivial infrared information: under transformations acting nonlinearly on the full correlator, the pole content retained within a given effective domain can change, as occurs explicitly under $SL(2,\mathbb{Z})$ transformations \cite{Amoretti:2026SL2Z}. Integrating out these modes exactly generally produces constitutive relations that are non-local in time and exhibit memory effects. The resultant non-local Green's functions generally have an exact embedding in Gaussian Schwinger-Keldysh field theory (we shall review this point in the article).}

{\ Maxwell-Cattaneo theory localises the retarded correlator by adding an auxiliary vector field that carries an isolated pole from the generic Green's function. In essence, in addition to charge transport, there is a spatial vector degree of freedom $v^{\mu}$ coupled to the charge current $J^{\mu}$ that at lowest order satisfies an equation of the form
	\begin{eqnarray}
		\label{Eq:SchematicMC}
		v^{i} + \tau \partial_{t} v^{i} = \ldots
	\end{eqnarray}
where $\tau$ is a relaxation time, $t,i$ are time and space coordinates and the ellipsis indicates spatial derivatives and higher derivative terms. It is the simplest local effective description of this physics, obtained by introducing a finite relaxation time and thereby promoting the leading non-hydrodynamic pole to an explicit dynamical degree of freedom. Such descriptions arise naturally in extended irreversible thermodynamics, relativistic M\"{u}ller--Israel--Stewart theories \cite{Romatschke:2009im,Montenegro:2016gjq}, holographic models \cite{Grozdanov:2018fic,Amoretti:2019cef,Baggioli:2023tlc,Liu:2024tqe,Ahn:2025mch}, and more generally as systematic truncations of the Mori-Zwanzig projection formalism \cite{teVrugt:2020,Amoretti:2023hpb}, where eliminating microscopic degrees of freedom produces memory kernels whose pole expansion may be represented by additional relaxation fields.A recent kinetic-theory analysis further showed the systematic emergence of Maxwell–Cattaneo-type dynamics in the effective slow sector when fast and slow relaxation scales are parametrically separated \cite{gavassino2026effectivefieldtheoryquasihydrodynamics}. Understanding the Schwinger-Keldysh embedding of Maxwell-Cattaneo theory is therefore a first step towards a systematic stochastic description of effective theories containing arbitrary collections of localised non-hydrodynamic poles and their continuum limits \cite{Amoretti:2026branchcuts}.}

{\ Rather than constructing hydrodynamics from a Schwinger-Keldysh action, as is typical in the literature, we shall address the inverse problem. Given a non-linear Maxwell-Cattaneo theory, when does there exist a Gaussian Schwinger-Keldysh embedding reproducing the prescribed constitutive relations? If such an embedding exists, is it unique? Answering these questions requires understanding not only the deterministic hydrodynamics, but also the associated entropy current, thermal noise and dynamical KMS symmetry.}

{\ Our starting point is the following observation. Suppose one has an external vector field $\zeta^{\mu}$ in addition to a $U(1)$ charge current $J^{\mu}$. Given a hydrostatic generating functional \cite{Banerjee:2012iz,Jensen:2012jh}, one knows precisely the hydrostatic part of the corresponding constitutive relations. The generating functional is constructed to exhaust all tensor structures compatible with symmetries that survive at hydrostaticity. Consequently, the hydrostatic constitutive relations are completely determined, while the remaining terms are proportional to thermal derivatives and are constrained by positivity of entropy production. Moreover, the hydrostatic generating functional captures all allowed improvements of the entropy current.}

{\ Suppose now that the additional vector field is instead intrinsically dissipative rather than external. What this means is that instead of satisfying the usual hydrostatic condition $\mathcal{L}_{\beta}\zeta^{\mu}=0$, we require
\begin{eqnarray}
    \zeta^{\mu}=0
\end{eqnarray}
at hydrostaticity. This ensures that the additional field leaves the static sector of the theory, and in particular the static Green's functions, unchanged. The hydrostatic generating functional therefore contains no dependence on $\zeta^{\mu}$, while tensor structures involving $\zeta^{\mu}$ become part of the non-hydrostatic constitutive relations. As a consequence, each such tensor structure acquires an independent transport coefficient unconstrained by hydrostaticity. Physically, local equilibrium does not depend on the additional degree of freedom $\zeta$; but when the fluid is moving $\zeta^{\mu}$ is non-zero from one local equilibrium patch to another and contributes to the flow.}

{\ More importantly for our analysis here, the entropy current improvements that do not vanish when $\mathcal{L}_{\beta}(*)=0$ are no longer completely determined by the hydrostatic generating functional. Once the dissipative vector is removed from the hydrostatic sector, additional entropy-current improvements built from $\zeta^{\mu}$ become possible. These improvements shift the constitutive relations entering the entropy production operator while leaving the hydrostatic sector unchanged. Consequently, unlike the conventional case, specifying the hydrostatic generating functional alone is insufficient to define the hydrodynamic theory: the entropy-current improvements must also be supplied independently.}

{\ The closest construction in the literature is that of \cite{Jain:2023obu}, where stable and causal Schwinger-Keldysh effective theories inspired by Maxwell-Cattaneo diffusion and M\"{u}ller-Israel-Stewart hydrodynamics were built directly, and where it was observed that such theories admit multiple inequivalent realisations of the dynamical KMS symmetry because ultraviolet sectors are not fixed by the infrared symmetries. Our inverse formulation makes this freedom quantitative: we determine, order by order in derivatives, which transport coefficients are constrained by the existence of a local Gaussian embedding and which remain genuine ambiguities of the enlarged description. Related Schwinger-Keldysh treatments of relaxational and quasi-diffusive charge transport include \cite{Grozdanov:2018fic,Abbasi:2022aao,Baggioli:2023tlc,Abbasi:2025qde}, while stochastic formulations of Israel-Stewart-type theories based on the information current have been developed in \cite{Mullins:2023tjg,Mullins:2023ott}. An alternative mechanism generating the same gapped pole, in which the additional vector originates from a weakly broken symmetry, has been developed in \cite{Armas:2021vku,Hongo:2024vbj}; there the hydrostatic sector is modified, in contrast with the intrinsically dissipative vector considered here, which leaves the static sector untouched.}

{\ In this paper we develop this construction explicitly for non-linear Maxwell-Cattaneo hydrodynamics. We first derive constitutive relations consistent with $PT$ symmetry and positivity of entropy production through second order in derivatives, identifying the corresponding entropy-current improvements when fluid velocity $u^{\mu}$ and temperature $T$ are frozen out. We then try to construct the embedding problem in Gaussian Schwinger-Keldysh theories realising the modified KMS transformation of \cite{Jain:2023obu}. For the class of embeddings we develop we show that the modified transformation restricts the transport coefficients but reproduces the desired form of the linearised Green's functions.}

\section{Non-linear Maxwell-Cattaneo quasi-hydrodynamics }

{\noindent In these sections we derive the generic non-linear Maxwell-Cattaneo equations for electric charge transport up to order two in derivatives using standard hydrodynamic procedures. In the next section we shall embed these expressions -with transport coefficient restrictions- into the Schwinger-Keldysh formalism and extract the corresponding stochastic noise.}

{\ Let us discuss field content before discussing symmetries. In addition to the usual charge and stress-energy-momentum tensor degrees of freedom, we have included the four-vector $\zeta^{\mu}$, at order one in derivatives so that it vanishes at equilibrium. As we work about a stationary background whose timelike Killing vector is aligned with the fluid velocity $u^{\mu}$, normalised so that $u^{\mu} u_{\mu} = -1$, we can decompose $\zeta^{\mu}$ into components
	\begin{eqnarray}
		\zeta^{\mu} &=& - \varphi u^{\mu} + v^{\mu} \; , \qquad u^{\mu} v_{\mu} = 0 \; , \qquad \Pi^{\mu \nu} = \eta^{\mu \nu} + u^{\mu} u^{\nu} \; , \qquad \Pi^{\mu \nu} v_{\nu} = v^{\mu} \; , \qquad  
	\end{eqnarray}
The spatial part of $\zeta^{\mu}$, given by $\Pi^{\mu \nu} \zeta_{\nu}$, shall eventually be identified with the field $v^{\mu}$ appearing in the Maxwell-Cattaneo equation \eqref{Eq:SchematicMC}.}

\subsection{Symmetries and entropy positivity } \label{sec:symandent}

{\ Let us now consider the relevant Ward identities that are satisfied by any theory with diffeomorphism and $U(1)$ gauge invariance. These take the schematic form
	\begin{subequations}
	\label{Eq:WardIDs}
	\begin{eqnarray}
		\label{Eq:WardID1}
		\nabla_{\mu} T^{\mu \nu} &=& F^{\nu \mu} J_{\mu} + \frac{1}{\sqrt{-g}} \frac{\delta S}{\delta \zeta_{\mu}} \nabla^{\nu} \zeta_{\mu} - \nabla_{\mu} \left( \frac{1}{\sqrt{-g}}  \frac{\delta S}{\delta \zeta_{\mu}} \zeta^{\nu} \right)  + (\mathrm{e.o.m}) \; , 	\\
		\nabla_{\mu} J^{\mu} &=& 0 + (\mathrm{e.o.m}) \; ,
	\end{eqnarray}
	\end{subequations}
where $(\mathrm{e.o.m})$ is present to acknowledge that Ward identities are valid off-shell up to satisfaction of the equations of motion of the fields of the theory. The term $\frac{1}{\sqrt{-g}} \frac{\delta S}{\delta \zeta_{\nu}}$ in \eqref{Eq:WardID1} is a response operator if $\zeta_{\mu}$ is external, while if $\zeta_{\mu}$ is internal this response operator is exactly the equation of motion for the vector field $\zeta_{\mu}$ and vanishes on-shell.}

{\ Let us then define the following quantity
	\begin{eqnarray}
		S_{\mathrm{can.}}^{\mu} = \frac{1}{T} \left[ p(\mu,T) u^{\mu} - T^{\mu \nu} u_{\nu} - \mu J^{\mu} - (u^{\nu} \zeta_{\nu}) V^{\mu}  \right] \; , \qquad V^{\mu} = \frac{1}{\sqrt{-g}}  \frac{\delta S}{\delta \zeta_{\mu}}  \; ,
	\end{eqnarray}
which is the canonical entropy current where $\mu$ and $T$ are two spacetime-dependent scalar fields. The total derivative of the canonical entropy current gives
	\begin{eqnarray}
		\label{Eq:DerivCanonS}
		\nabla_{\mu} S_{\mathrm{can.}}^{\mu} &=& \nabla_{\mu} \left( \frac{p(\mu,T)}{T} u^{\mu} \right) - V^{\mu} (\mathcal{L}_{\beta} \zeta)_{\mu} 
						    -  T^{\mu \nu} \nabla_{(\mu} \left( \frac{u_{\nu)}}{T} \right) \nonumber \\
						    &\;& + \frac{1}{T} J^{\mu} \left( E_{\mu} - T \nabla_{\mu} \left( \frac{\mu}{T} \right) \right) 
						    	  + (\mathrm{e.o.m}) \; ,
	\end{eqnarray}
after employing the Ward identities \eqref{Eq:WardIDs}. In the standard approach to hydrodynamics we are instructed to isolate the non-hydrostatic pieces of the currents, i.e. those which vanish upon imposing that the Lie derivative of any terms vanishes $\mathcal{L}_{\beta}(*)$ along $\beta^{\mu}$ the thermal-vector\footnote{In the case of the gauge field one must generalise $\mathcal{L}_{\beta}(*)$ to $\delta_{\beta}$ which allows for a gauge transformation., and develop an entropy current improvement term of the form
	\begin{eqnarray}
		\label{Eq:HydrostaticS}
		\nabla_{\mu} S_{\mathrm{HS}}^{\mu} &=& -  \nabla_{\mu} \left( \frac{p(\mu,T)}{T} u^{\mu} \right) + V_{\mathrm{HS}} (\mathcal{L}_{\beta} \zeta)_{\mu}+ T_{\mathrm{HS}}^{\mu \nu} \nabla_{(\mu} \left( \frac{u_{\nu)}}{T} \right) \nonumber \\
						    &\;& - \frac{1}{T} J_{\mathrm{HS}}^{\mu} \left( E_{\mu} - T \nabla_{\mu} \left( \frac{\mu}{T} \right) \right)  \; . 
	\end{eqnarray}
The construction is always assumed to be exhaustive, in that every possible term compatible with \eqref{Eq:HydrostaticS} and any symmetries at hydrostaticity is included. Notice that this identity is an off-shell relationship and thus lacks the term $(\mathrm{e.o.m})$ compared to \eqref{Eq:DerivCanonS}.}

{\ Adding the improvement term \eqref{Eq:HydrostaticS} to our expression for the divergence of the canonical current \eqref{Eq:DerivCanonS} leads to the expression 
	\begin{subequations}
	\begin{eqnarray}
			\label{Eq:DivergenceS}
			\nabla_{\mu} S^{\mu}
		&=& \nabla_{\mu} \left( S_{\mathrm{can.}}^{\mu} + S_{\mathrm{HS}}^{\mu} \right) \nonumber \\
		&=& - \left( V^{\mu} - V^{\mu}_{\mathrm{HS}} \right) (\mathcal{L}_{\beta} \zeta)_{\mu} 
			 - \frac{1}{2} \left( T^{\mu \nu} - T^{\mu \nu}_{\mathrm{HS}} \right)  (\mathcal{L}_{\beta} g)_{\mu \nu} 
			 + \left( J^{\mu} - J^{\mu}_{\mathrm{HS}} \right) (\delta_{\beta} B)_{\mu} \nonumber \\
		&\;& + (\mathrm{e.o.m})  \; , \qquad
	\end{eqnarray}
where 
	\begin{eqnarray}
		\label{Eq:EntropyDefs}
		(\delta_{\beta} B)_{\mu} &=& \mathcal{L}_{\beta} B_{\mu} + \delta_{\Lambda} B_{\mu} = \left( E_{\mu} - T \nabla_{\mu} \left( \frac{\mu}{T} \right) \right) \; , \qquad B_{\mu} = A_{\mu} + \partial_{\mu} \phi \; , \\
		\mathcal{L}_{\beta} g_{\mu \nu} &=& 2 \nabla_{(\mu} \beta_{\nu)} \; , 
	\end{eqnarray}
	\end{subequations}
with $A_{\mu}$ an external gauge field, $\phi$ a phase field and $\delta_{\Lambda}$ a gauge transformation. We have made these definitions to make explicit connections with our later Schwinger-Keldysh action, and to indicate the role of the thermal derivative in the divergence of the entropy current \eqref{Eq:DivergenceS}.}

{\ For an external vector field $\zeta^{\mu}$, at this point the story is complete in form. Given a hydrostatic generating functional, which is always constructed to exhaust all potential terms that can contribute to the hydrostatic constitutive relations, we then enumerate all terms proportional to the thermal derivatives in the constitutive relations. Requiring that $\nabla_{\mu} S^{\mu} \geq 0$ then imposes constraints only on this latter set of (dissipative and non-hydrostatic/non-dissipative) terms.}

{\ However, in the case of the Maxwell-Cattaneo equations, the ``hydrostatic condition'' is not just $D \zeta^{\mu}=0$ but instead $\zeta^{\mu}=0$. Thus $\zeta^{\mu}$ does not appear in the hydrostatic generating functional and, as stated above, this ensures that the new vector field does not affect the static sector of the Green's functions. Consequently,
	\begin{eqnarray}
		V^{\mu}_{\mathrm{HS}} \equiv 0 \; ,
	\end{eqnarray}
independently of satisfaction of the equations of motion. Each tensor structure in the constitutive relations proportional to $\zeta^{\mu}$, and non-vanishing upon imposing $\mathcal{L}_{\beta}(*)=0$, now is promoted to have its own independent transport coefficient in a manner compatible with symmetries because they have joined the non-hydrostatic sector. The divergence of the entropy current off-shell then takes the schematic form
		\begin{eqnarray}
			\label{Eq:DivergenceS2}
			\nabla_{\mu} S^{\mu}
		&=& \nabla_{\mu} \left( S_{\mathrm{can.}}^{\mu} + S_{\mathrm{HS}}^{\mu} \right) \nonumber \\
		&=& - V^{\mu}  (\mathcal{L}_{\beta} \zeta)_{\mu}  
			 - \frac{1}{2} \left( T^{\mu \nu} - T^{\mu \nu}_{\mathrm{HS}} \right) (\mathcal{L}_{\beta} g)_{\mu \nu}
			 + \left( J^{\mu} - J^{\mu}_{\mathrm{HS}} \right) (\delta_{\beta} B)_{\mu} \nonumber \\
		&\;& + (\mathrm{e.o.m})  \; .
	\end{eqnarray}
However, in liberating $\zeta_{\mu}$ from the hydrostatic sector, we have no longer constructed the most general entropy current improvement term because we have not accounted for improvements that can be built from $\zeta^{\mu}$. We can readily identify such terms from the hydrostatic generating functional because in that case the hydrostatic sector does exhaust all possibilities for improvement not proportional to $\mathcal{L}_{\beta}(*)$ i.e.
	\begin{align}
		 & (J^{\mu},V^{\mu},T^{\mu \nu})_{\mathrm{HS}} - \left[  (J^{\mu},V^{\mu},T^{\mu \nu})_{\mathrm{HS}} \right]_{\zeta=0} & (\mathrm{external \; \zeta}) \nonumber \\
		\rightarrow & (\Delta J^{\mu},\Delta V^{\mu},\Delta T^{\mu \nu})  & (\mathrm{internal \; \zeta}) \; . 
	\end{align}
Thus the divergence of entropy current becomes
	\begin{eqnarray}
			\label{Eq:DivergenceComplete}
			\nabla_{\mu} S^{\mu}
		&=& \nabla_{\mu} \left( S_{\mathrm{can.}}^{\mu} + S_{\mathrm{HS}}^{\mu} + \Delta S^{\mu} \right) \nonumber \\
		&=& - ( V^{\mu} - \Delta V^{\mu} ) (\mathcal{L}_{\beta} \zeta)_{\mu} 
			 - \frac{1}{2} \left( T^{\mu \nu} - T^{\mu \nu}_{\mathrm{HS}} - \Delta T^{\mu \nu} \right) (\mathcal{L}_{\beta} g)_{\mu \nu} \nonumber \\
		&\;& + \left( J^{\mu} - J^{\mu}_{\mathrm{HS}} - \Delta J^{\mu} \right) (\delta_{\beta} B)_{\mu}
			+ (\mathrm{e.o.m})  \; .
	\end{eqnarray}
In particular, taking the divergence of the entropy current on-shell and demanding it be positive definite we have
	\begin{eqnarray}
			\label{Eq:DivergenceCompleteOnshell}
			\nabla_{\mu} S^{\mu}
		&=& \Delta V^{\mu} (\mathcal{L}_{\beta} \zeta)_{\mu} 
			 - \frac{1}{2} \left( T^{\mu \nu} - T^{\mu \nu}_{\mathrm{HS}} - \Delta T^{\mu \nu} \right) (\mathcal{L}_{\beta} g)_{\mu \nu} \nonumber \\
		&\;& + \left( J^{\mu} - J^{\mu}_{\mathrm{HS}} - \Delta J^{\mu} \right) (\delta_{\beta} B)_{\mu}  \geq 0 \; .
	\end{eqnarray}	
In the external field case, the hydrostatic generating functional fixed the definition of $S^{\mu}$ in terms of coefficients appearing in the functional. Now we must supply not only the hydrostatic generating functional, but the form of the improvements to specify $S^{\mu}$. We have also identified a new class of terms which do not vanish when $\mathcal{L}_{\beta}(*)=0$, so they are non-dissipative in that sense, but they do vanish at hydrostaticity. We shall explore this in more detail with explicit examples at order one and order two in derivatives.}

{\ Having established the general entropy-current analysis, we may set $u^\mu$, $T$ and $\varphi$ to be constants and proceed to construct the constitutive relations in a convenient operator basis. However, we note that when energy-momentum fluctuations are retained, the status of $u^\mu$ and $T$ away from equilibrium involves additional hydrodynamic-frame ambiguities. A recent line of work has advocated interpreting these ambiguities as a gauge-like reparametrisation redundancy of fluctuating hydrodynamics \cite{Dore:2021xqq,Sampaio:2025jtp,Torrieri:2026fjz}. We avoid this issue here by restricting to the fixed-$u^\mu$, fixed-$T$ charge sector.}

\subsection{Constitutive relations, equations of motion and constraints }

{\noindent In the next paragraph we shall write down the equations of motion for the Maxwell-Cattaneo system to order two in derivatives. To do this however, we need to further delineate our operator basis. We could of course write down all possible terms, however this basis is necessarily redundant as many terms are related to each other by lower order equations of motion. We shall always choose to use lower order equations to ensure that there is only a single time derivative acting on the hydrodynamic fields $\zeta^{\mu}$ and $\mu$ in the equations of motion. This does not mean that background fields may not contain higher time derivatives, nor that we have completely fixed our frame by making this choice as we shall discuss. When necessary we shall refer to this choice as the Langevin canonical form for the hydrodynamic equations.}

{\ The scalar part $\varphi = u^{\mu} \zeta_{\mu}$ of $\zeta_{\mu}$ can, and following the principles of effective theory should, be non-zero in a general basis. In our Langevin canonical form, the resultant scalar equation of motion would have at most one time derivative. We shall choose for $D \varphi \sim \mathcal{O}(\partial^2)$ and always assume that there is an order one term in the scalar equation of the form $\varphi$. Consequently, we can always eliminate the time derivative of $\varphi$ and make the resultant equation, $u_{\mu} V^{\mu}=0$ algebraic. This should be compared to the equation $\Pi_{\mu \nu} V^{\nu}=0$ which gives the dynamics for $v^{\mu} \sim \mathcal{O}(\partial)$. There we will treat $v^{\mu}$, $\tau D v^{\mu}$ as $\sim \mathcal{O}(\partial)$ which introduces a pole that cannot be eliminated and reproduces an equation of the form \eqref{Eq:SchematicMC}. We note that while one could also choose for the scalar to be given independent dynamics just like the spatial part, this leads to the existence of a gapped pole that vanishes from the spectrum for homogeneous perturbations and is not our primary interest.}

{\ With our operator basis chosen, we can define the following hydrostatic generating functional and isolate the hydrostatic terms in the constitutive relations at order two in derivatives. This generating functional has the form
	\begin{subequations}
	\label{Eq:HydrostaticG}
	\begin{eqnarray}
		W &=& \int d^{3+1}x \; \sqrt{-g} \left( P(\mu) + \frac{1}{2} \chi_{E} E^2 
				 + \frac{1}{2} \chi_{B} \mathcal{B}^{2} + \mathcal{O}(\partial^4) \right) \; , \qquad \\
			p^{\mu}
		&=& \chi_{E}  E^{\mu}  + \mathcal{O}(\partial^3)  \; ,  \qquad m^{\mu} 
		= \chi_{B} \mathcal{B}^{\mu} \; , \\
			E^{\mu}
		&=& F^{\mu \nu} u_{\nu} \; , \qquad \mathcal{B}^{\mu} = \frac{1}{2 } \epsilon^{\mu \nu \rho \sigma} u_{\nu} F_{\rho \sigma} \; , 
	\end{eqnarray}
	\end{subequations}
where all coefficients are functions of $\mu$ and we have used the hydrostaticity condition $E^{\mu} - \partial_{\perp}^{\mu} \mu=0$ to eliminate any derivatives of $\mu$. To all orders in derivatives \cite{Kovtun:2016lfw} one finds
	\begin{eqnarray}
		J_{\mathrm{HS}}^{\mu} &=& n u^{\mu} - \nabla_{\nu} M^{\mu \nu} \; , \qquad M^{\mu \nu} = p^{\mu} u^{\nu} - p^{\nu} u^{\mu} - \epsilon^{\mu \nu \rho \sigma} u_{\rho} m_{\sigma} \; .
	\end{eqnarray}
Consequently,
	\begin{eqnarray}
		 J^{\mu}_{\mathrm{HS}} &=& \left( n + \chi_{E} \nabla_{\nu} E^{\nu} + \frac{\partial \chi_{E}}{\partial \mu} E^{\nu} \partial^{\perp}_{\nu} \mu \right) u^{\mu}  - \chi_{E} D E^{\mu} - \frac{\partial \chi_{E}}{\partial \mu} (D\mu) E^{\mu}  \nonumber \\
		 &\;& +\ \chi_{B} \epsilon^{\mu \nu \rho \sigma} u_{\rho} \nabla_{\nu} \mathcal{B}_{\sigma} 
		 + \frac{\partial \chi_{B}}{\partial \mu} \epsilon^{\mu \nu \rho \sigma} u_{\rho}  ( \nabla_{\nu}^{\perp} \mu ) \mathcal{B}_{\sigma} + \mathcal{O}(\partial^3) \; . 
	\end{eqnarray}
We remind the reader that $V^{\mu}_{\mathrm{HS}}\equiv 0$.}

{\ Adding in dissipative terms, into which $\zeta^{\mu}$ is included, the most generic non-linear equations of motion to order two in derivatives for our fields given in Langevin canonical form are
	\begin{subequations}
	\label{Eq:EquationsofMotion}
	\begin{eqnarray}
		\label{Eq:Current}
			v^{\mu}  + \tau D v^{\mu} + \frac{r \tau}{\alpha} \partial^{\mu}_{\perp} \mu  
		&+& \theta_{2} \epsilon^{\mu \nu \rho \sigma} u_{\nu} (\partial_{\rho}^{\perp} \mu) \mathcal{B}_{\sigma} \nonumber \\
		&=& \frac{r \tau}{\alpha} E^{\mu} +  \theta_{1} D E^{\mu}  + \epsilon^{\mu \nu \rho \sigma} u_{\nu} \left( \theta_{2} E_{\rho} + \theta_{3} v_{\rho}   \right) \mathcal{B}_{\sigma} \nonumber \\
	&\;&+ \mathcal{O}(\partial^3) \; , \qquad \\
		\label{Eq:Charge}
		   	\chi_{\rho \rho} D \mu + \alpha \partial^{\perp}_{\mu} v^{\mu}  + \frac{\partial \alpha}{\partial \mu} v^{\mu} \partial_{\mu}^{\perp} \mu &-& (\sigma_{\mathrm{DC}} - r \tau)  \partial^{2}_{\perp} \mu  - \frac{\partial (\sigma_{\mathrm{DC}} - r \tau)}{\partial \mu} (\partial_{\perp} \mu)^2
			    \nonumber \\
			   &=& - (\sigma_{\mathrm{DC}} - r \tau) \partial_{\mu}^{\perp} E^{\mu}  - \frac{\partial (\sigma_{\mathrm{DC}} - r \tau)}{\partial \mu} E^{2}   \nonumber \\
			   &\;& + \frac{\partial (\sigma_{\mathrm{DC}} - r \tau)}{\partial \mu} E^{\mu} ( E_{\mu} - \partial_{\mu}^{\perp} \mu )  + \mathcal{O}(\partial^3) \; , \qquad 
	\end{eqnarray}
		\end{subequations}
where the $U(1)$ charge current is given by 
	\begin{eqnarray}
		\label{Eq:ChargeCurrentConstitutive}
		J^{\mu} &=& \left[ n + \mathfrak{b}_{0} v^{\nu} ( E_{\nu} - \partial_{\nu}^{\perp} \mu ) + \chi_{E} \partial_{\nu}^{\perp} E^{\nu} + \mathfrak{c}_{1} \left( \partial_{\nu}^{\perp} E^{\nu} - \partial_{\perp}^2 \mu \right) 
		+ \mathfrak{b}_{1} \partial_{\nu}^{\perp} v^{\nu} +  \frac{\partial \chi_{E}}{\partial \mu} E^2   \right. \nonumber \\
	&\;& \left. + ( \mathfrak{b}_{5} + \frac{\partial \chi_{E}}{\partial \mu} ) E^{\nu} ( E_{\nu} - \partial_{\nu}^{\perp} \mu ) + \mathfrak{b}_{2} E^{\nu} v_{\nu}  + \mathfrak{b}_{6} ( E - \partial^{\perp} \mu )^2 + \mathfrak{b}_{3} v^{2}  \right. \nonumber \\
	&\;& \left.  
		+ \frac{1}{2} \frac{\partial \chi_{B}}{\partial \mu} \mathcal{B}^{2} \right] u^{\mu} 
		+ (\sigma_{\mathrm{DC}} - r \tau) (E^{\mu} - \partial_{\perp}^{\mu} \mu) + \alpha v^{\mu}  
		- ( \chi_{E} - \mathfrak{c}_{2} ) D E^{\mu}
		  \nonumber \\
	&\;& + \epsilon^{\mu \nu \rho \sigma} u_{\nu} \left( \chi_{B} \partial^{\perp}_{\rho} + \frac{\partial \chi_{B}}{\partial \mu}  E_{\rho} 
		 + \left( \mathfrak{c}_{3}  - \frac{\partial \chi_{B}}{\partial \mu}  \right) ( E_{\rho} - \partial_{\rho}^{\perp} \mu)   + \mathfrak{b}_{4} v_{\rho} \right) \mathcal{B}_{\sigma} \nonumber \\
	&\;& + \mathcal{O}(\partial^3) \; . 
	\end{eqnarray}
Terms on the left-hand-side of the equalities in \eqref{Eq:EquationsofMotion} are what remain when the background fields are tuned to zero. Conversely, terms on the right hand side represent how the fluid couples to the background. We have used \eqref{Eq:Current} to eliminate any potential $Dv^{\mu}$ term from \eqref{Eq:ChargeCurrentConstitutive}.}	

{\ We have rendered the equations of motion in Langevin canonical form \eqref{Eq:EquationsofMotion}, which is easy at this order. At the next order we can see that time derivatives of \eqref{Eq:ChargeCurrentConstitutive} will introduce terms of the form $D v^{\mu}$ from the time component into the equation of motion \eqref{Eq:Charge}. It is not just enough to eliminate additional time derivatives from the constitutive relations; time derivatives in the equations of motion must also be eliminated using the analogue of \eqref{Eq:Current} at the appropriate derivative order to reduce to our desired format.}

{\ As we discussed, eliminating time derivatives does not entirely fix frame. To maintain the Langevin canonical form we cannot make frame transformations that involve the hydrodynamic fields $\mu$ and $v^{\mu}$. However, we can make frame transformations that only involve the background fields, for example:
	\begin{subequations}
	\label{Eq:FrameTrans}
	\begin{eqnarray}
		\label{Eq:vmuframe}
		v^{\mu} &\rightarrow& \tilde{v}^{\mu} = v^{\mu} + c_{1} D E^{\mu} + \epsilon^{\mu \nu \rho \sigma} u_{\nu} \left( c_{2} E_{\rho} + c_{3} v_{\rho}   \right) \mathcal{B}_{\sigma} + \mathcal{O}(\partial^3) \; , \qquad \\
		\mu &\rightarrow& \tilde{\mu} = \mu + c_{4} \partial_{\mu} E^{\mu} + c_{5} E^{2} + c_{6} \mathcal{B}^{2} + \mathcal{O}(\partial^3) \; . 
	\end{eqnarray}
	\end{subequations}
We note for the particular case of $v^{\mu}$, frame transformations that lead to terms that do not vanish at hydrostaticity in \eqref{Eq:Current} are not permitted, which is why there is no term of the form $E^{\mu}$ in \eqref{Eq:vmuframe}. Nevertheless, these frame transformations \eqref{Eq:FrameTrans} reshuffle the background couplings in our Langevin canonical form equations \eqref{Eq:EquationsofMotion} and we can frame fix \eqref{Eq:EquationsofMotion} by transforming
	\begin{eqnarray}
		\label{Eq:FrameChoice}
		v^{\mu} &\rightarrow& v^{\mu} -   \theta_{1} D E^{\mu} -  \epsilon^{\mu \nu \rho \sigma} u_{\nu} \left( \theta_{2} ( E_{\rho} - \partial_{\rho}^{\perp} \mu)  + \theta_{3} v_{\rho}   \right) \mathcal{B}_{\sigma} + \mathcal{O}(\partial^3) \; .
	\end{eqnarray}
Thus we can take $\theta_{1}=\theta_{2}=\theta_{3}=0$ in \eqref{Eq:EquationsofMotion} with other coefficients re-defined suitably. In fact, it is clear that this can be done at any order in derivatives so that \eqref{Eq:Current} will only receive corrections order by order that explicitly involve $v^{\mu}$ or $E^{\mu} - \partial_{\perp}^{\mu} \mu$ such as the following operators
	\begin{eqnarray}
		E^2 v^{\mu}\; , \; \; \mathcal{B}^{2} (E^{\mu} - \partial_{\perp}^{\mu} \mu) \; , \; \; v^2 v^{\mu} \; , \; \; \mathrm{etc.}
	\end{eqnarray}
at third order in derivatives.}

{\ Before leaving our non-linear analysis here, we comment on formally integrating out the auxiliary field $v^{\mu}$. In particular, in our frame choice up to order two in derivatives we can formally invert \eqref{Eq:Current}. Define the relaxation operator
\begin{equation}
\label{Eq:RelaxationOperator}
 \mathbbm{K}^\mu{}_\nu
 =
 \delta^\mu{}_\nu+\tau D\,\delta^\mu{}_\nu \; .
\end{equation}
which we assume is invertible so that we can subsequently solve \eqref{Eq:Current} and determine $v^{\mu}$ to be
\begin{equation}
 v^\mu
 =   \frac{r\tau}{\alpha}
 (\mathbbm{K}^{-1})^\mu{}_\nu \bigl(E^\nu-\partial_\perp^\nu\mu\bigr) \; .
\end{equation}
The $U(1)$ current, after substituting in this result, then becomes
\begin{eqnarray}
\label{Eq:Nonlocalconstitutive}
	J^{\mu} &=& \left[ n + \mathfrak{b}_{0} v^{\nu} ( E_{\nu} - \partial_{\nu}^{\perp} \mu ) + \chi_{E} \partial_{\rho}^{\perp} E^{\rho} + \left( \mathfrak{c}_{1} \partial_{\rho}^{\perp} + \mathfrak{b}_{1} \partial_{\nu}^{\perp} (\mathbbm{K}^{-1})\indices{^{\nu}_{\rho}}  \right) \left( E^{\rho} - \partial_{\perp}^{\rho} \mu \right)   \right. \nonumber \\
	&\;& \left. +  \frac{\partial \chi_{E}}{\partial \mu} E^2 + E_{\alpha} \left( ( \mathfrak{b}_{5} + \frac{\partial \chi_{E}}{\partial \mu} ) \Pi^{\alpha \nu} + \mathfrak{b}_{2} (\mathbbm{K}^{-1})^{\alpha \nu} \right) ( E_{\nu} - \partial_{\nu}^{\perp} \mu )   \right. \nonumber \\
	&\;& \left.  + ( E_{\nu} - \partial^{\perp}_{\nu} \mu ) ( \mathfrak{b}_{6} \Pi^{\nu \rho}+ \mathfrak{b}_{3} (\mathbbm{K}^{-2})^{\nu \rho} ) ( E_{\rho} - \partial^{\perp}_{\rho} \mu ) 
		+ \frac{1}{2} \frac{\partial \chi_{B}}{\partial \mu} \mathcal{B}^{2} \right] u^{\mu} \nonumber \\
	&\;& + \left( (\sigma_{\mathrm{DC}} - r \tau) \Pi^{\mu \nu} + \alpha  (\mathbbm{K}^{-1})^{\mu \nu} \right) (E_{\nu} - \partial^{\perp}_{\nu} \mu) 
		- ( \chi_{E} - \mathfrak{c}_{2} ) D E^{\mu}
		  \nonumber \\
	&\;& + \epsilon^{\mu \nu \rho \sigma} u_{\nu} \left( \chi_{B} \partial^{\perp}_{\rho} + \frac{\partial \chi_{B}}{\partial \mu}  E_{\rho} 
		 + \left( \left( \mathfrak{c}_{3}  - \frac{\partial \chi_{B}}{\partial \mu} \right) \Pi_{\rho}^{\alpha} + \mathfrak{b}_{4} (\mathbbm{K}^{-1})_{\rho}^{\alpha} \right) ( E_{\alpha} - \partial_{\alpha}^{\perp} \mu) \right) \mathcal{B}_{\sigma} \nonumber \\
	&\;& + \mathcal{O}(\partial^3) \; ,
\end{eqnarray}
where $\mathbbm{K}^{-2} = \mathbbm{K}^{-1} \cdot \mathbbm{K}^{-1}$. Thus an explicit non-locality has been introduced into the constitutive relation through $\mathbbm{K}$. We shall see this at the level of the linearised equations in the next section, and revisit it after we define the Schwinger-Keldysh action and when we discuss stochastic quasi-hydrodynamics in sections \ref{Sec:IntegratingOutAdditionalField} and \ref{sec:stochastic} respectively.}

\subsubsection{Linearised analysis }\label{sec:linearised}

{\noindent To confirm interpretations of terms in \eqref{Eq:EquationsofMotion}, such as our identification of the DC conductivity, let us compute the linearised retarded Green's functions about a background with non-zero chemical potential and vanishing electromagnetic field by varying the background gauge field. The linearised equations take the form
	\begin{subequations}
	\label{Eq:LinEquationsofMotion}
	\begin{eqnarray}
		\label{Eq:LinCurrent}
			\delta v^{\mu}  + \tau D \delta v^{\mu} + \frac{r \tau}{\alpha} \partial^{\mu}_{\perp} \delta \mu  
		&=& \frac{r \tau}{\alpha} \delta E^{\mu} + \mathcal{O}(\partial^3,\delta^2) \; , \qquad \\
		   	\chi_{\rho \rho} D \delta \mu + \alpha \partial^{\perp}_{\mu} \delta v^{\mu}  &-& (\sigma_{\mathrm{DC}} - r \tau)  \partial^{2}_{\perp} \delta \mu \nonumber \\
			   &=&- (\sigma_{\mathrm{DC}} - r \tau) \partial_{\mu}^{\perp} \delta E^{\mu}  + \mathcal{O}(\partial^3,\delta^2) \; , \qquad 
	\end{eqnarray}
where the $U(1)$ charge current is given by
	\begin{eqnarray}
		\label{Eq:LinChargeCurrentConstitutive}
		\delta J^{\mu} &=& \left[ \chi_{\rho \rho} \delta \mu + \chi_{E} \partial_{\nu}^{\perp} \delta E^{\nu} + \mathfrak{c}_{1} \left( \partial_{\nu}^{\perp} \delta E^{\nu} - \partial_{\perp}^2 \delta \mu \right) 
		+ \mathfrak{b}_{1} \partial_{\nu}^{\perp} \delta v^{\nu}  \right] u^{\mu} \nonumber \\
	&\;&  + (\sigma_{\mathrm{DC}} - r \tau) (\delta E^{\mu} - \partial_{\perp}^{\mu} \delta \mu) + \alpha \delta v^{\mu}  
		- ( \chi_{E} -  \mathfrak{c}_{2}) D \delta E^{\mu}  
		  \nonumber \\
	&\;&  + \chi_{B}  \epsilon^{\mu \nu \rho \sigma} u_{\nu} \partial^{\perp}_{\rho} \delta \mathcal{B}_{\sigma} 
		+ \mathcal{O}(\partial^3) \; . 
	\end{eqnarray}
			\end{subequations}
At this point it is useful to compare with the formalism in \cite{Amoretti:2025kem} as the linearised analysis of our system was performed exhaustively in that work. The variable $\delta \bar{J}^{\mu}$ in their formalism is given by
	\begin{eqnarray}
		\label{Eq:Barjdef}
		\delta \bar{J}^{\mu} &=& \alpha \delta v^{\mu}  
		+ (\sigma_{\mathrm{DC}} - r \tau) (\delta E^{\mu} - \partial_{\perp}^{\mu} \delta \mu)  
		  + \mathfrak{c}_{2} D \delta E^{\mu} + \mathcal{O}(\partial^3) \; , \qquad 
	\end{eqnarray}
so that $U(1)$ charge conservation becomes
	\begin{eqnarray}
		\label{Eq:Barjchargecons}
		\chi_{\rho \rho} D \mu + \partial_{\mu}^{\perp} \delta \bar{J}^{\mu} = 0 \; . 
	\end{eqnarray}
We can solve for $v^{\mu}$ from \eqref{Eq:Barjdef} and substitute into \eqref{Eq:LinCurrent}. This introduces time derivatives of $\mu$ which must be eliminated using \eqref{Eq:Barjchargecons} to bring us to the Langevin canonical form. We find
	\begin{eqnarray}
		\label{Eq:Barjextrafield}
		\delta \bar{J}^{\mu} + \tau D \delta \bar{J}^{\mu} - \sigma_{\mathrm{DC}} ( \delta E^{\mu} - \partial^{\mu}_{\perp} \delta \mu  )
		&=& \left( \mathfrak{c}_{2} + \tau (\sigma_{\mathrm{DC}} - r \tau) \right) D \delta E^{\mu} \nonumber \\
		&\;& + \mathcal{O}(\partial^3,\delta^2)
	\end{eqnarray}
Equations \eqref{Eq:Barjchargecons} and \eqref{Eq:Barjextrafield} are the analogues of the linearised equations of \cite{Amoretti:2025kem}.}

{\ Having confirmed that the linearised equations can be brought into the form used in \cite{Amoretti:2025kem}, we now compute the charge conductivity rather than the individual current-current correlators. We introduce the following transform
	\begin{eqnarray}
		f(t,\vec{x}) = \int \frac{d^{3+1}k}{(2\pi)^4} \; f(\omega,\vec{k}) e^{-i ( \omega t - \vec{k} \cdot \vec{x} )} 
	\end{eqnarray}
for any function $f(t,\vec{x})$, and subsequently take the wave-vector $\vec{k}$ to lie along the $x$-direction. The longitudinal conductivity and transverse (to the wavevector) conductivities are defined by
\begin{subequations}
\begin{eqnarray}
    \sigma_{(\mathrm{L})}(\omega,k)
    &=& \frac{1}{-i\omega}\left[
    \left\langle J^{x}J^{x}\right\rangle_{\mathrm{R}}(\omega,k)
    -\left\langle J^{x}J^{x}\right\rangle_{\mathrm{R}}(0,k)
    \right] \; , \\
    \sigma_{(\mathrm{T})}(\omega,k)
    &=& \frac{1}{-i\omega}\left[
    \left\langle J^{y}J^{y}\right\rangle_{\mathrm{R}}(\omega,k)
    -\left\langle J^{y}J^{y}\right\rangle_{\mathrm{R}}(0,k)
    \right] \; .
\end{eqnarray}
\end{subequations}
This is the one-hydrodynamic-pole and one-gapped-pole specialisation of the Mittag-Leffler decomposition of \cite{Amoretti:2025kem}. In the present theory it is useful to introduce
\begin{eqnarray}
    \Xi(\omega,k)
    &\equiv& -i\omega\chi_{\rho\rho}(1-i\omega\tau)
    +k^{2} \left( \sigma_{\mathrm{DC}} - i \omega \tau ( \sigma_{\mathrm{DC}}-r\tau) \right) \; .
    \label{Eq:LinearResponseDenominator}
\end{eqnarray}
Solving \eqref{Eq:LinEquationsofMotion} and substituting into \eqref{Eq:LinChargeCurrentConstitutive} gives
\begin{subequations}
\label{Eq:CompactConductivities}
\begin{eqnarray}	
	\label{Eq:CompactConductivitiesL}
    \sigma_{(\mathrm{L})}(\omega,k)
    &=&  - \frac{i\omega\chi_{\rho\rho} \left( \sigma_{\mathrm{DC}} - i \omega \tau ( \sigma_{\mathrm{DC}}-r\tau) \right) }{\Xi(\omega,k)}
    + i\omega\left(\chi_{E}-\mathfrak{c}_{2}\right)
    +\mathcal{O}(\partial^{3}) \; , \qquad \\
    \label{Eq:CompactConductivitiesT}
     \sigma_{(\mathrm{T})}(\omega,k)
    &=&
 \frac{\sigma_{\mathrm{DC}} - i\omega\tau (\sigma_{\mathrm{DC}}-r\tau )}{1-i\omega\tau}
    + i \omega ( \chi_E- \mathfrak{c}_{2} )
    +O(\partial^3) \; .
\end{eqnarray}
\end{subequations}
The polynomial in frequency term of \eqref{Eq:CompactConductivitiesL} is the polarisation contribution. The denominator in the first term of \eqref{Eq:CompactConductivitiesL} contains both the hydrodynamic $\omega_{\mathfrak{D}}$ and gapped mode $\omega_{\mathrm{gap}}$, while \eqref{Eq:CompactConductivitiesT} contains only the transverse mode $\omega_{\mathrm{T}}$ which are defined by 
\begin{subequations}
\label{Eq:LinearResponsePolesandResidues}
\begin{align}
    \omega_{\mathfrak{D}}(k)
    &=
    -i\frac{\sigma_{\mathrm{DC}}}{\chi_{\rho\rho}}k^2
    +O(k^4),
    &\qquad
    \mathcal{R}_{\mathfrak{D}}(k)
    &=
    -\frac{\sigma_{\mathrm{DC}}^2}{\chi_{\rho\rho}}k^2
    +O(k^4) \; ,
    \\
    \label{Eq:LinearisedGappedPole}
    \omega_{\mathrm{gap}}(k)
    &=
    -\frac{i}{\tau}
    +i\frac{r\tau}{\chi_{\rho\rho}}k^2
    +O(k^4),
    &\qquad
    \mathcal{R}_{\mathrm{gap}}(k)
    &=
    r
    +\frac{r\tau
        \bigl(2\sigma_{\mathrm{DC}}-r\tau\bigr)}
        {\chi_{\rho\rho}}k^2
    +O(k^4) \; , \\
    \omega_{\mathrm{T}}
    &=
    -\frac{i}{\tau},
    &\qquad
    \mathcal{R}_{\mathrm{T}}
    &= r \; .
\end{align}
\end{subequations}
Here $\mathcal{R}$ represents the residue of the corresponding pole. We note that the residue of the diffusive pole vanishes as $k\rightarrow0$, whereas the gapped pole remains visible in the homogeneous AC conductivity. Moreover, the Einstein relation read from the hydrodynamic pole is
\begin{eqnarray}
    \mathfrak{D}=\frac{\sigma_{\mathrm{DC}}}{\chi_{\rho\rho}} \; ,
\end{eqnarray}
as expected for a gapped theory consisting solely of charge transport.}

{\ Equation \eqref{Eq:CompactConductivities} can be expanded into the explicit Mittag--Leffler form of \cite{Amoretti:2025kem} using the expressions in \eqref{Eq:LinearResponsePolesandResidues}:
\begin{subequations}
\begin{eqnarray}
    \sigma_{(\mathrm{L})}(\omega,k)
    &=& \underbrace{\frac{i\mathcal{R}_{\mathfrak{D}}(k)}
    {\omega-\omega_{\mathfrak{D}}(k)}}_{\text{hydrodynamic pole}}
    +\underbrace{\frac{i\mathcal{R}_{\mathrm{gap}}(k)}
    {\omega-\omega_{\mathrm{gap}}(k)}}_{\text{gapped pole}} \nonumber \\
    &\;& +\underbrace{(\sigma_{\mathrm{DC}} - r \tau)  +i\omega\left(\chi_{E}-\mathfrak{c}_{2}\right)
    +\mathcal{O}(\omega^{2})}_{\text{holomorphic terms}} \; , \\
    \sigma_{(\mathrm{T})}(\omega,k)
    &=& \underbrace{\frac{i\mathcal{R}_{\mathrm{T}}}
    {\omega-\omega_{\mathrm{T}}}}_{\text{transverse pole}}
    +\underbrace{(\sigma_{\mathrm{DC}} - r \tau) +i\omega\left(\chi_{E}-\mathfrak{c}_{2}\right) +\mathcal{O}(\omega^{2})}_{\text{holomorphic terms}} \; .
\end{eqnarray}
\end{subequations}
In particular, in the homogeneous limit
\begin{subequations}
\begin{eqnarray}
    &\;& \sigma_{(\mathrm{L})}(\omega,0) 
    = \sigma_{(\mathrm{T})}(\omega,0)
    = \frac{i r}{\omega+i/\tau} +
    \left(\sigma_{\mathrm{DC}}-r\tau\right) +i\omega\left(\chi_{E}-\mathfrak{c}_{2}\right)
    +\mathcal{O}(\omega^{2}) \; , \qquad \\
    &\;& \lim_{\omega\rightarrow0} \sigma_{(\mathrm{L})}(\omega,0) 
    = \lim_{\omega\rightarrow0} \sigma_{(\mathrm{T})}(\omega,0) = \sigma_{\mathrm{DC}} \; .
\end{eqnarray}
\end{subequations}
The constant $\sigma_{\mathrm{DC}}-r\tau$ and the polarisation term proportional to $\chi_{E}-\mathfrak{c}_{2}$ are precisely the holomorphic data left at this derivative order when we can truncate a full thermal theory to a diffusive and gapped pole. These terms can represent contributions from other poles or branch cuts that are not relevant in the frequency regime we consider. In particular, to compare with \cite{Jain:2023obu}, we note that if one chooses $r \tau = \sigma_{\mathrm{DC}}$, so that the response is dominated by a single gapped pole then we recover the standard Drude-style conductivity in the homogeneous limit
	\begin{eqnarray}
	 	&\;& \sigma_{(\mathrm{L})}(\omega,0) 
    = \sigma_{(\mathrm{T})}(\omega,0)
    = \frac{\sigma_{\mathrm{DC}}}{1- i \tau \omega} +i\omega \chi_{E} +\mathcal{O}(\omega^{2})
	\end{eqnarray}
The formalism we have developed allows us to capture low frequency and wave-vector information coming from higher poles that have already passed out of the low frequency regime. These higher-order operators generally correct the holomorphic part order by order in $i\omega$ and $k^{2}$, while additional non-hydrodynamic poles must be added explicitly rather than generated perturbatively (see \cite{Amoretti:2025kem} for details).}

\subsubsection{Entropy positivity at $\mathcal{O}(\partial)$ }\label{sec:orderoneentropy}

{\noindent Let us now consider constraints on the transport coefficients imposed by positivity of entropy production. We will need to solve for the evolution of the $v^{\mu}$. In our reduced operator basis this is easy to do and we find
	\begin{eqnarray}
		\label{Eq:vmueqnO1}
		v^{\mu} &=& - \tau D v^{\mu} + \frac{r \tau}{\alpha} (E^{\mu} - \partial_{\perp}^{\mu} \mu) +  \mathcal{O}(\partial^2) \; . 
	\end{eqnarray}
Suppose now that we try to impose positivity of entropy production on the transport terms without adding improvement terms. Taking the entropy current on-shell we find
	\begin{eqnarray}
		T \nabla_{\mu} S^{\mu} &=&  \left( (\sigma_{\mathrm{DC}} - r \tau) (E^{\mu} - \partial_{\perp}^{\mu} \mu) + \alpha v^{\mu} \right) (E_{\mu} - \partial_{\mu}^{\perp} \mu) \geq 0 \; . 
	\end{eqnarray}
Using the equation of motion \eqref{Eq:vmueqnO1} to eliminate $v^{\mu}$ it subsequently follows that
	\begin{eqnarray}
		T \nabla_{\mu} S^{\mu} &=& (\sigma_{\mathrm{DC}} - r \tau)  \left( (E - \partial_{\perp} \mu) + \frac{\alpha v}{2 (\sigma_{\mathrm{DC}} - r \tau)} \right)^2 \nonumber \\
		&\;& - \frac{\alpha^2 v^2}{4 (\sigma_{\mathrm{DC}} - r \tau)}   \geq 0 \; . 
	\end{eqnarray}
There is no way to make both coefficients of the quadratic terms positive definite unless $\alpha=0$. Inspecting \eqref{Eq:EquationsofMotion} we see that for the equations to remain finite when $\alpha \rightarrow 0$, we require that $r\tau/\alpha = \mathrm{constant}$ and thus at least one of $\tau$ or $r$ must go to zero. The former would force an algebraic equation for $v^{\mu}$, so let us assume that $r \rightarrow 0$. In the linearised analysis of section \ref{sec:linearised} we see that the gapped pole we were attempting to add drops from the spectrum when $r \rightarrow 0$, see \eqref{Eq:LinearisedGappedPole}, defeating the purpose of introducing the auxiliary field $v^{\mu}$. Without improvement, we cannot achieve our aim of localising the additional pole and have positivity of entropy production.}

{\ Thus we turn to the hydrostatic generating functional to obtain suitable improvement terms. To $\mathcal{O}(\partial^3)$ in derivatives it has the form
		\begin{subequations}
		\label{Eq:HydrostaticO1}
	\begin{eqnarray}
		W_{<\mathcal{O}(\partial^3)} &=& \int d^{3+1}x \; \sqrt{-g} \left( P + \frac{\chi_{E}}{2} E^2 + \frac{\chi_{B}}{2} \mathcal{B}^{2} + \rho_{\varphi}^{(1)} \varphi  + \frac{\rho_{\varphi}^{(2)}}{2} \varphi^2 - \frac{\rho_{v}}{2} v^2 + f_{1} ( v \cdot E)  \right. \nonumber \\
		&\;& \left. \hphantom{ \int d^{3+1}x \; \sqrt{-g} \left( \right.} \vphantom{\frac{f_{3}}{2}} + f_{2} \nabla_{\perp} \cdot E  + f_{3} \nabla_{\perp} \cdot v +  \mathcal{O}(\partial^3) \right) \; , \\
			V^\mu_{\mathrm{ext,HS}}
		&=&  \left( \rho_{\varphi}^{(1)} + \rho_{\varphi}^{(2)} \varphi \right) u^{\mu} 
			 + \left( - \rho_{v} v^{\mu} + f_{1} E^{\mu} - \frac{\partial f_{3}}{\partial \mu} \nabla_{\perp}^{\mu} \mu  \right) + \mathcal{O}(\partial^2), \\
			p^{\mu}
		&=& \chi_{E} E^{\mu} +  f_{1} v^{\mu} - \frac{\partial f_{2}}{\partial \mu} \nabla_{\perp}^{\mu} \mu \; , \qquad m^{\mu} 
		= \chi_{B} \mathcal{B}^{\mu} \; .
	\end{eqnarray}
	\end{subequations}
Thus, in line with the discussion of section \ref{sec:symandent}, we find the maximal non-hydrostatic improvement terms to be
	\begin{subequations}
	\begin{eqnarray}
		 \Delta V^{\mu}
		&=& (\rho_{\varphi}^{(1)} + \rho_{\varphi}^{(2)} \varphi ) u^{\mu} - \rho_{v} v^{\mu} + \left( f_{1} - \frac{\partial f_{3}}{\partial \mu} \right)  E^{\mu} + \frac{\partial f_{3}}{\partial \mu} (E^{\mu} - \nabla_{\perp}^{\mu} \mu) \nonumber \\
		&\;& + \mathcal{O}(\partial^2) \; , \qquad \\
		\Delta J^{\mu} &=& 0 + \mathcal{O}(\partial^2)  \; .
	\end{eqnarray}
	\end{subequations}
These terms we use to redefine the entropy current as in \eqref{Eq:DivergenceComplete} so that it can support our Maxwell-Cattaneo mode non-trivially. Again, now that we have discussed the general functional, we will drop the $\varphi$ sector.}

{\ With the $\Delta$ terms in hand we find that \eqref{Eq:DivergenceCompleteOnshell} becomes
	\begin{eqnarray}
		T \nabla_{\mu} S^{\mu} &=& - \left( \rho_{v} v^{\mu}  - \left( f_{1} - \frac{\partial f_{3}}{\partial \mu} \right)  E^{\mu} -\frac{\partial f_{3}}{\partial \mu} (E^{\mu} - \nabla_{\perp}^{\mu} \mu) \right) D v_{\mu} \nonumber \\
						    &\;& +  \left( (\sigma_{\mathrm{DC}} - r \tau) (E^{\mu} - \partial_{\perp}^{\mu} \mu) + \alpha v^{\mu} \right) (E_{\mu} - \partial_{\mu}^{\perp} \mu)  \geq 0 + \mathcal{O}(\partial^3) \; . \qquad
	\end{eqnarray}
Note that the improvement terms can be written as a total derivative which vanishes at hydrostaticity (i.e. when $\zeta_{\mu}=0$) of the form
	\begin{eqnarray}
		\Delta S^{\mu} &=& \left( \frac{\rho_{v}}{2} v^2 - \left( f_{1} -\frac{\partial f_{3}}{\partial \mu} \right) E  \cdot v  + \frac{\partial f_{3}}{\partial \mu}  ( E - \partial_{\perp} \mu)  \cdot v \right) u^{\mu} + \mathcal{O}(\partial^3) \; ,
	\end{eqnarray}
where we remind the reader that $v^{\mu}, Dv^{\mu} \sim \mathcal{O}(\partial)$. The first term is precisely of the form generated by adding a piece $\chi_{v} \mathrm{d} v^2$ to the differential form of the first law in \cite{Jain:2023obu}, however our interpretation is clearly different. In \cite{Jain:2023obu} the authors modify the first law to include an energy contribution of this form - however, in equilibrium such a contribution is always zero and it cannot affect the thermodynamics. Instead, we interpret it as a missing improvement one can make to the entropy which vanishes in equilibrium and at hydrostaticity.}

{\ It is slightly more interesting for interpretation to replace $Dv^{\mu}$ using \eqref{Eq:vmueqnO1} in the positivity condition at $\mathcal{O}(\partial)$. Completing the square gives the rather complicated expression
\begin{eqnarray}
\label{Eq:FullEntropy}
T\nabla_{\mu}S^{\mu}
&=& \frac{\rho_{v}}{\tau}
\left[
 v^{\mu}
 +\frac{\tau(\alpha^{2}-\rho_{v}r)-\alpha\,\frac{\partial f_{3}}{\partial\mu}}
 {2\alpha\rho_{v}}
 (E^{\mu}-\partial_{\perp}^{\mu}\mu)
 -\frac{f_{1}-\frac{\partial f_{3}}{\partial\mu}}
 {2\rho_{v}}E^{\mu}
\right]^{2}
\nonumber\\
&&+\left[
 \sigma_{\mathrm{DC}}-r\tau
 +\frac{r}{\alpha}\frac{\partial f_{3}}{\partial\mu}
 -\frac{\tau}{4\rho_{v}}
 \left(
  \frac{\alpha^{2}-\rho_{v}r}{\alpha}
  -\frac{1}{\tau}\frac{\partial f_{3}}{\partial\mu}
 \right)^{2}
\right]
\nonumber\\
&&\quad\times\left[
 (E^{\mu}-\partial_{\perp}^{\mu}\mu)
 +\frac{f_{1}-\frac{\partial f_{3}}{\partial\mu}}
 {2\left[
 \sigma_{\mathrm{DC}}-r\tau
 +\frac{r}{\alpha}\frac{\partial f_{3}}{\partial\mu}
 -\frac{\tau}{4\rho_{v}}
 \left(
  \frac{\alpha^{2}-\rho_{v}r}{\alpha}
  -\frac{1}{\tau}\frac{\partial f_{3}}{\partial\mu}
 \right)^{2}
 \right]}
\right.
\nonumber\\
&&\left.\qquad\times\left(
 \frac{r}{\alpha}
 +\frac{1}{2\rho_{v}}
 \left[
  \frac{\alpha^{2}-\rho_{v}r}{\alpha}
  -\frac{1}{\tau}\frac{\partial f_{3}}{\partial\mu}
 \right]
\right)E^{\mu}
\right]^{2}
\nonumber\\
&&-\frac{1}{4}
\left(f_{1}-\frac{\partial f_{3}}{\partial\mu}\right)^{2}
\left\{
 \frac{1}{\rho_{v}\tau}
 +\frac{
 \left(
  \frac{r}{\alpha}
  +\frac{1}{2\rho_{v}}
  \left[
   \frac{\alpha^{2}-\rho_{v}r}{\alpha}
   -\frac{1}{\tau}\frac{\partial f_{3}}{\partial\mu}
  \right]
 \right)^{2}}
 {\sigma_{\mathrm{DC}}-r\tau
 +\frac{r}{\alpha}\frac{\partial f_{3}}{\partial\mu}
 -\frac{\tau}{4\rho_{v}}
 \left(
  \frac{\alpha^{2}-\rho_{v}r}{\alpha}
  -\frac{1}{\tau}\frac{\partial f_{3}}{\partial\mu}
 \right)^{2}}
\right\}E^{2} \nonumber \\
&\;& +\mathcal{O}(\partial^{3}) \; .
\label{Eq:EntropyCompletedSquares}
\end{eqnarray}
We see immediately if $\tau>0$ we need $\rho_{v}>0$. Moreover, the coefficient of the second square term in \eqref{Eq:FullEntropy} must be positive which we can see as a constraint on the DC conductivity. On the other hand, the coefficient of the final $E^{2}$ square is always non-positive and thus the coefficient must be zero which requires
\begin{eqnarray}
\label{Eq:f1f3hydrostaticity}
 f_{1}=\frac{\partial f_{3}}{\partial\mu} \; .
\end{eqnarray}
Substituting this expression into the hydrostatic generating functional of \eqref{Eq:HydrostaticO1} then shows that both $f_{1}$ and $f_{3}$ can be taken to be zero because, after integration by parts, the remaining term after satisfying \eqref{Eq:f1f3hydrostaticity} is proportional to a hydrostaticity condition.}

{\ After imposing the relation of \eqref{Eq:f1f3hydrostaticity}, positivity is then equivalent to demanding that $\rho_{v}>0$ and
\begin{subequations}
\label{Eq:SteadyState}
\begin{eqnarray}
\label{Eq:PositivityBound}
 \sigma_{\mathrm{DC}}\geq
 \frac{\tau 
 (\alpha^{2}+\rho_{v}r)^{2}}
 {4\alpha^{2}\rho_{v}} \; .
\end{eqnarray}
This latter constraint replaces the usual positivity constraint on $\sigma_{\mathrm{DC}}$. We note in passing that if the bound in \eqref{Eq:PositivityBound} is saturated, entropy production can vanish without imposing $E-\partial_{\perp}\mu=0$; instead one finds
\begin{eqnarray}
 v^{\mu}=\frac{
 \tau(\rho_{v}r - \alpha^{2})}
 {2\alpha\rho_{v}}
 (E^{\mu}-\partial_{\perp}^{\mu}\mu) \; .
\end{eqnarray}
\end{subequations}
This is reminiscent of many electrically driven systems \cite{Amoretti:2022ovc,Brattan_2024,Amoretti:2024jig}, and the additional constraints required for this locus to solve the remaining hydrodynamic equations are worthy of development elsewhere.}

\subsubsection{Entropy positivity at $\mathcal{O}(\partial^2)$ }

{\noindent Moving onto the next order, the hydrostaticity generating functional for an external vector field has the form
	\begin{eqnarray}
		W_{<\mathcal{O}(\partial^4)} &=& W_{<\mathcal{O}(\partial^3)} + \int d^{3+1}x \; \sqrt{-g} \left( \epsilon^{\mu \nu \rho \sigma} u_{\nu} ( \chi_{v}^{(1)} v_{\mu} E_{\rho}  + \chi_{v}^{(2)} \nabla_{\rho} v_{\mu} ) \mathcal{B}_{\sigma} \right. \nonumber \\
		&\;& \left. \hphantom{W_{<\mathcal{O}(\partial^3)} + \int d^{3+1}x \; \sqrt{-g} \left(  \right.} + \mathcal{O}(\partial^4) \right) \; .
	\end{eqnarray}
Thus the improvement terms for the entropy current become
	\begin{subequations}
\begin{align}
\Delta V^\mu
={}&-\rho_v v^\mu
+\left(
  \chi_v^{(1)}
 -\frac{\partial\chi_v^{(2)}}{\partial\mu}
 \right)
 \epsilon^{\mu\nu\rho\sigma}
 u_\nu E_\rho \mathcal{B}_\sigma
\nonumber\\
&-\chi_v^{(2)}
 \epsilon^{\mu\nu\rho\sigma}
 u_\nu\nabla_\rho \mathcal{B}_\sigma
+ \frac{\partial\chi_v^{(2)}}{\partial\mu}
 \epsilon^{\mu\nu\rho\sigma}
 u_\nu
 \left(E_\rho - \nabla^\perp_\rho\mu\right)\mathcal{B}_\sigma
+O(\partial^3) \;  , \\
\Delta J^\mu
={}& - u^\mu\left( \frac{1}{2}\frac{\partial\rho_v}{\partial\mu}v^2
\right)
 +O(\partial^3) \; . 
\end{align}
	\end{subequations}
Employing our frame choice of \eqref{Eq:FrameChoice}, we find that the divergence of the entropy current can be written as
\begin{subequations}
\label{eq:entropy-divergence-squares}
\begin{align}
T\nabla_\mu S^\mu
={}&
\frac{\rho_v}{\tau}
\left[
v^\mu
+\frac{\tau\left(\alpha^2-\rho_v r\right)}
       {2\alpha\rho_v}
\left(E^\mu-\partial_\perp^\mu\mu\right)
-\frac{Q^\mu}{2\rho_v} \right. \nonumber \\
& \left. \hphantom{\frac{\rho_v}{\tau} \left[ \right.} +\frac{\tau}{2\rho_v} \left(\mathfrak{b}_{4}
+\frac{1}{\tau} \frac{\partial\chi_v^{(2)}}{\partial\mu}  \right) \epsilon^{\alpha\nu\mu\sigma}
u_\nu
\left(E_\alpha-\partial^\perp_\alpha\mu\right)
\mathcal{B}_\sigma 
\right]^2
\nonumber \\
&\quad
+ \left( \sigma_{\mathrm{DC}}
-\frac{\tau\left(\alpha^2+\rho_v r\right)^2}
       {4\alpha^2\rho_v} \right)
\left[
E^\mu-\partial_\perp^\mu\mu
+\frac{R^\mu}{2\left(\sigma_{\mathrm{DC}}
-\frac{\tau\left(\alpha^2+\rho_v r\right)^2}
       {4\alpha^2\rho_v}\right)}
\right]^2 \nonumber \\
& +O(\partial^4) \; ,
\end{align}
where the new tensor structures are
\begin{align}
Q^\mu
&\equiv
\left(
\chi_v^{(1)}
-\frac{\partial\chi_v^{(2)}}{\partial\mu}
\right)
\epsilon^{\mu\nu\rho\sigma}
u_\nu E_\rho \mathcal{B}_\sigma
-\chi_v^{(2)}
\epsilon^{\mu\nu\rho\sigma}
u_\nu\partial_\rho \mathcal{B}_\sigma \; ,
\\  
R^\mu
&\equiv
\mathfrak{c}_2\,D E^\mu
+\frac{\alpha^2+\rho_v r}
       {2\alpha\rho_v}\,Q^\mu ,
\end{align}
\end{subequations}
and the quartic derivative has been used to absorb a term of the form $R^2$. We see immediately that the conditions on being positive definite are unchanged at this order in derivatives from the order one case discussed in section \ref{sec:orderoneentropy}. However, the naive steady state condition we previously discussed \eqref{Eq:SteadyState} has become subtle requiring that $R^{\mu}=0$. This is unsurprising however as we expect electrically driven steady states to generically require fine tuning for time varying electric fields and/or terms that seek to induce vortical motion.}

\section{Embedding as Maxwell-Cattaneo in Schwinger-Keldysh }

{\noindent Thus far we have worked with equations of motion for a $U(1)$ current and additional dissipative vector field, seeking to constrain them to have an additional gapped pole, Onsager reciprocity and positivity of entropy production. The construction of a Schwinger--Keldysh effective action is a stronger inverse problem than the construction of hydrodynamic equations. Constitutive relations and entropy production constrain the deterministic evolution, but an action must additionally encode a consistent off-shell symmetry realization, the fluctuation sector, and positivity of the path-integral measure. This distinction is familiar from ordinary mechanics: equations of motion may contain non-conservative forces that cannot be obtained from a conventional local action without enlarging the system or introducing nonlocality.

{\ In the present setting, the observable linearised theory could be written directly as a nonlocal Schwinger-Keldysh theory, which we will show. The point of introducing the Maxwell-Cattaneo auxiliary field is that it localises the theory. In this section, we demonstrate that a local Gaussian action which is invariant under the modified KMS symmetry of \cite{Jain:2023obu} can reproduce a subset of the hydrodynamic equations developed earlier through second order. The local action formulation seems to be more constrained than the underlying constitutive theory. The reason is clear: requiring a dynamical KMS involution to act on the auxiliary field in a prescribed manner imposes additional relations among the transport coefficients that are not implied by entropy positivity or by the observable current correlators.}

{\ We interpret these additional relations as constraints on the local Schwinger-Keldysh embedding rather than necessarily on the underlying hydrodynamics. Just as the existence of an action principle is an additional property of a system of equations, the existence of a local, manifestly positive and KMS-invariant stochastic action is additional structure beyond the second law. Whether the generic hydrodynamic formulation of the previous section can be embedded generally remains an open question.}

\subsection{Review of KMS constraints with $U(1)$ charge and an external vector field }\label{sec:standardKMS}

{\noindent There are two conceptually distinct questions we seek to clarify by reviewing the standard KMS construction. First, what is the relation between dissipation and Gaussian fluctuations? Second, how is this relation modified by genuinely non-Gaussian stochastic processes? To answer this, we begin with the most general local SK action and examine the constraints imposed by the Glorioso-Liu formulation of dynamical KMS \cite{Glorioso:2017fpd}. This is the backbone of formulating the inverse problem in later sections.}

{\ We begin with the usual symmetry doubling procedure \cite{Haehl:2014zda,Haehl:2015uoc,Haehl:2016pec,Crossley:2015evo,Glorioso:2017fpd,Glorioso:2016gsa,Gao:2017bqf,Haehl:2018lcu,Glorioso:2018wxw,Jensen:2018hse} with $_{a}$ subscripts appended to conjugate response field terms. In particular, let us reintroduce
	\begin{eqnarray}
		(B_{,a})_{\mu} &=& (A_{,a})_{\mu} + \partial_{\mu} \phi_{,a} \; ,
	\end{eqnarray}
where $B_{\mu}$ was originally defined in \eqref{Eq:EntropyDefs}. Suppose we take $\zeta_{\mu}$ to be an external field for the moment and that we write down the most generic action which closes under KMS symmetry,
		\begin{subequations}
				\label{Eq:KMStrans}
		\begin{eqnarray}
		(B_{a}(x),\zeta_{a}(x))_{\mu} &\rightarrow& \Theta \left[ (B_{a}(\vartheta x),\zeta_{a}(\vartheta x))_{\mu} - i ( \mathcal{L}_{\beta} B(\vartheta x), \mathcal{L}_{\beta}\zeta(\vartheta x) )_{\mu} \right] \; , \\
		(B(x),\zeta(x))_{\mu} &\rightarrow& \Theta \left[ (B(\vartheta x),\zeta(\vartheta x))_{\mu} \right] \; , 
	\end{eqnarray}
	\end{subequations}
to the order in derivatives that we wish to work. Here $\vartheta$ reverses the time coordinate and $\Theta$ is the time reversal operator acting on the fields\footnote{The full table of PT assignments is given in appendix \ref{appendix:PTtable}. We also ensure invariance under the time-independent diagonal shift symmetry
\begin{displaymath}
    B_{r,\mu}\longrightarrow B_{r,\mu} +\partial_\mu \lambda(\vec{x})\;, \qquad B_{a\mu} \longrightarrow B_{a\mu}\;,
\end{displaymath}
so that we do not have spontaneous $U(1)$ symmetry breaking~\cite{Firat:2025upx}.}. Having done this, one can subsequently isolate the linear and quadratic parts in the response fields to arrive at an action of the form
	\begin{eqnarray}
		\label{Eq:SKAction}
		S &=& \int d^{3+1} x \;  \left[ X_{a}^{T} \left( \left( \begin{array}{c} J \\ V \end{array} \right) +
		 i T \mathbbm{N} X_{a}  \right)  + \mathcal{O}(X_{a}^3) \right] \; , \qquad X_{a} =  \left( \begin{array}{c} B_{a} \\ \zeta_{a} \end{array} \right)  \;, \qquad 
	\end{eqnarray}
where the objects $J$ and $V$ are arbitrary functions of $B$ and $\zeta$ and their derivatives while $\mathbbm{N}$ is a matrix operator, dependent on the same fields, acting to the right. We note that because of the contraction with the response field and its transpose, we can take $\mathbbm{N}$ to be self-adjoint and shall do so henceforth. Varying the action \eqref{Eq:SKAction} with respect to $X_{a}$ and subsequently setting $X_{a}=0$ gives $(J,V)$, the deterministic constitutive relations. We can always decompose these objects according to
	\begin{eqnarray}
    \label{Eq:RealSpaceConstitutive}
    \left(\begin{array}{c} 
    J \\
    V
    \end{array}\right) = \left(\begin{array}{c} 
    J \\
    V
    \end{array}\right)_{\mathrm{HS}}  +  \mathbbm{D} \left( \begin{array}{c} 
    \mathcal{L}_\beta B_r \\
    \mathcal{L}_\beta \zeta_r
    \end{array} \right) \; .
\end{eqnarray}
where $_{\mathrm{HS}}$ indicates terms that are non-zero at hydrostaticity and $\mathbbm{D}$ is an operator acting to the right.}

{\ We will need to define the adjoint of a differential operator. With respect to the spacetime inner product given by
\begin{equation}
    \langle X,Y\rangle
    \equiv
    \int d^{d+1}x\,\sqrt{-g}\,X_I Y^I ,
\end{equation}
where $X$ and $Y$ are two fields, the adjoint of an operator $O$ satisfies
\begin{equation}
    \langle X,\mathcal OY\rangle
    =
    \langle \mathcal O^\dagger X,Y\rangle ,
\end{equation}
where boundary terms in the integral are assumed to vanish. Thus, we can further decompose $\mathbbm{D}$ into it's self-adjoint and anti-self-adjoint pieces
	\begin{eqnarray}
		\label{Eq:DissipativeDeconstruction}
		\mathbbm{D} &=& \mathbbm{D}_{S} +  \mathbbm{D}_{A} \; , \qquad \mathbbm{D}_{S,A} =  \frac{1}{2} \left( \mathbbm{D} \pm \mathbbm{D}^{\dagger} \right) \; .
	\end{eqnarray}
Being self-adjoint the first term in \eqref{Eq:DissipativeDeconstruction} has real eigenvalues, while the second has imaginary eigenvalues; the self-adjoint term will correspond to the dissipative sector of hydrodynamics while the anti-self-adjoint term typically gives the non-dissipative, non-hydrostatic sector.}

{\  Under a KMS transformation \eqref{Eq:KMStrans}, the action \eqref{Eq:SKAction} becomes
	\begin{eqnarray}
		\label{Eq:SKActionTrans}
		&\;& \int d^{3+1} x \;  \left[ - i (\mathcal{L}_{\beta} X)^{T} \Theta^{T} \left( \begin{array}{c} J \\ V \end{array} \right)_{\mathrm{HS}}  + i ( \mathcal{L}_{\beta} X)^{T} ( \Theta^{T} \mathbbm{D}_{S} \Theta - T \Theta^{T} \mathbbm{N} \Theta) ( \mathcal{L}_{\beta} X)  \right. \nonumber \\
		&\;& \left. \hphantom{ + \int d^{3+1} x \;  \left[ \right.} \vphantom{ \left( \begin{array}{c} J \\ V \end{array} \right)_{\mathrm{HS}}} + X_{a} ^{T} \left( - ( \Theta^{T} \mathbbm{D}_{S} \Theta + \mathbbm{D}_{S} ) - ( \Theta^{T} \mathbbm{D}_{A} \Theta + \mathbbm{D}_{A}) + 2 T \Theta^{T} \mathbbm{N} \Theta \right) (\mathcal{L}_{\beta} X) \right. \nonumber \\
		&\;& \left. \hphantom{ + \int d^{3+1} x \;  \left[ \right.} \vphantom{ \left( \begin{array}{c} J \\ V \end{array} \right)_{\mathrm{HS}}} + X_{a}^{T} \left( \Theta^{T} \left( \begin{array}{c} J \\ V \end{array} \right)_{\mathrm{HS}}  - \left( \begin{array}{c} J \\ V \end{array} \right)_{\mathrm{HS}}  \right) + i T X_{a}^{T} \left( \Theta^{T} \mathbbm{N} \Theta - \mathbbm{N} \right) X_{a} \right. \nonumber \\
		&\;&  \left. \hphantom{ + \int d^{3+1} x \;  \left[ \right.} \vphantom{ \left( \begin{array}{c} J \\ V \end{array} \right)_{\mathrm{HS}}}  +  \left( \mathcal{K}  \mathcal{O}(X_{a}^3) - \mathcal{O}(X_{a}^3) \right) \right]
	\end{eqnarray}
where we have suppressed arguments of the operators above and $\mathcal{K}  \mathcal{O}(X_{a}^3)$ represents the KMS variation of order three and higher terms in the response fields $X_{a}$. If we have any chance to embed the hydrodynamic theory in a KMS complete action, then certain constraints that we can derive from this variation must hold. For example, upon imposing hydrostaticity it must be the case that
	\begin{eqnarray}
		\label{Eq:HSKMSconstraint}
		\Theta^{T} \left( \begin{array}{c} J[\Theta X] \\ V[\Theta X] \end{array} \right)_{\mathrm{HS}}  &=& \left( \begin{array}{c} J[X] \\ V[X] \end{array} \right)_{\mathrm{HS}} \; . 
	\end{eqnarray}
This constraint is independent of the higher response term contributions. For the next contributions we assume that either $ \left( \mathcal{K}  \mathcal{O}(X_{a}^3) - \mathcal{O}(X_{a}^3) \right)$ is KMS closed, by which we mean it vanishes, or it is at least of order $X_{a}^3$ so that we can derive the Onsager-Casimir relations:
	\begin{subequations}
	\label{Eq:OnsagerCasimir}
	\begin{eqnarray}
		\Theta^{T} \mathbbm{D}_{S}[\Theta X] \Theta &=& \mathbbm{D}_{S}[X] \; , \\
		\Theta^{T} \mathbbm{D}_{A}[\Theta X] \Theta &=& - \mathbbm{D}_{A}[X] \; , \\
		\Theta^{T} \mathbbm{N}[\Theta X] \Theta &=& \mathbbm{N}[X] \; .
	\end{eqnarray}
	\end{subequations}
It subsequently follows that
	\begin{eqnarray}
		\label{Eq:Intermediate1}
		&\;& \int d^{3+1} x \;  \left[ - i (\mathcal{L}_{\beta} X)^{T} \left( \begin{array}{c} J \\ V \end{array} \right)_{\mathrm{HS}}  + i ( \mathcal{L}_{\beta} X)^{T} ( \mathbbm{D}_{S} - T \mathbbm{N} ) ( \mathcal{L}_{\beta} X)  \right. \nonumber \\
		&\;& \left. \hphantom{ + \int d^{3+1} x \;  \left[ \right.} \vphantom{ \left( \begin{array}{c} J \\ V \end{array} \right)_{\mathrm{HS}}} + 2 X_{a} ^{T} \left( - \mathbbm{D}_{S}  + T \mathbbm{N} \right) (\mathcal{L}_{\beta} X)  +  \left( \mathcal{K}  \mathcal{O}(X_{a}^3) - \mathcal{O}(X_{a}^3) \right) \right] \; .
	\end{eqnarray}
The first term is a total derivative, by definition of the hydrostatic sector, while for the remaining two terms to vanish so that the action is KMS invariant we see that we must impose the usual relation encoding the fluctuation-dissipation theorem:
	\begin{eqnarray}
		\label{Eq:SimpleFDT}
		\mathbbm{D}_{S} = T \mathbbm{N} \; . 
	\end{eqnarray}
As $\mathbbm{D}_{S}$ is the term that appears in the condition for positivity of entropy production, we see that positivity of $\mathbbm{N}$ which is implied by unitarity of the action, also implies positivity of entropy production.}
	
{\ Given a supplied hydrodynamic theory it has a Gaussian KMS invariant action if \eqref{Eq:HSKMSconstraint} and \eqref{Eq:OnsagerCasimir} are satisfied by the hydrodynamics and one imposes \eqref{Eq:SimpleFDT}. The Gaussian sector of such a theory is, by definition, KMS closed. To this action one can additionally include any KMS closed polynomial that starts at $\mathcal{O}(X_{a}^3)$; for example 
	\begin{eqnarray}
		\mathcal{O}(X_{a}^3) = i \left( X_{a}^{T} M (X_{a} - i \mathcal{L}_{\beta} X) \right)^3 \; ,
	\end{eqnarray}
with $M$ a symmetric, real matrix satisfying $M[X] = \Theta^{T} M[\Theta X] \Theta$. The result is the same (non-linear) hydrodynamics with the same relationship between noise and dissipation. In appendix \ref{appendix:highernoise} we show how higher response terms affect the story when they are not KMS closed i.e. when $ \left( \mathcal{K}  \mathcal{O}(X_{a}^3) - \mathcal{O}(X_{a}^3) \right)$ does contain terms at order $X_{a}^{0}$ and $X_{a}^{1}$ .}

{\ To summarise the results of this section: if we are supplied a hydrodynamic theory which satisfies the Onsager-Casimir relations \eqref{Eq:OnsagerCasimir} then we can embed this theory in a Gaussian Schwinger-Keldysh action (i.e. quadratic in $X_a$) which is KMS invariant. Importantly, we do this by choosing the noise matrix to be given by \eqref{Eq:SimpleFDT}. We can then add higher response terms as we please while maintaining our hydrodynamic constitutive relations and positivity of entropy production on the condition that such terms are KMS invariant.}

\subsection{Embedding Maxwell-Cattaneo }

{\noindent The constraints developed from the embedding of hydrodynamics in the previous sections do not naively carry over to our Maxwell-Cattaneo system. In particular, vanishing of the variation relied on decomposing the constitutive relations as in \eqref{Eq:RealSpaceConstitutive}, i.e. into a hydrostatic part and a non-hydrostatic part proportional to the thermal derivatives of the fields. We have terms in our constitutive relations that fall into neither of these categories.}

{\ One way to resolve the tension was suggested in \cite{Jain:2023obu}, namely - modify the standard KMS transformation. In particular, $\zeta_\mu$ can be treated as the thermal derivative of a different vector field, which conjugate field is $\zeta_a$.  Otherwise, the $B_{a}$ and $B$ fields transform as usual. Thus, under a KMS transformation we have:
\begin{subequations}
				\label{Eq:KMStransModified}
		\begin{eqnarray}
		(B_{a}(x),\zeta_{a}(x))_{\mu} &\rightarrow& \bar{\Theta} \left[ (B_{a}(\vartheta x),\zeta_{a}(\vartheta x))_{\mu} - i ( \mathcal{L}_{\beta} B(\vartheta x), \zeta(\vartheta x) )_{\mu} \right] \; , \\
		(B(x),\zeta(x))_{\mu} &\rightarrow& \Theta \left[ (B(\vartheta x),\zeta(\vartheta x))_{\mu} \right] \; , \\
		\Theta = \mathbbm{1}_{2} &\;& \bar{\Theta} = \left( \begin{array}{cc} 1 & 0 \\ 0 & -1 \end{array} \right) \; .
	\end{eqnarray}
	\end{subequations}
We can combine these transformations into our $X_{a}$ vector as
	\begin{eqnarray}
		\label{Eq:ModifiedKMSX}
		X_{a} \rightarrow \bar{\Theta}[ X_{a} - i \mathcal{L} X ] \; , \qquad \mathcal{L} X &=& \left(\begin{array}{c} \mathcal{L}_{\beta} B \\ \zeta \end{array} \right) \; ,
	\end{eqnarray}
and, in a later section, we shall give a definite interpretation to \eqref{Eq:KMStransModified}. For now we note that the decomposition of \eqref{Eq:RealSpaceConstitutive} is replaced by
	\begin{eqnarray}
	\label{Eq:ModifiedRealSpaceConstitutive}
	    \left(\begin{array}{c} 
    J \\
    V
    \end{array}\right) = \left(\begin{array}{c} 
    J \\
    V
    \end{array}\right)_{\mathrm{HS}}  +  \mathbbm{D} \mathcal{L}X \; .
\end{eqnarray}
In particular, $\mathbbm{D}$ acts on $\zeta$, not $\mathcal{L}_{\beta} \zeta$. The generic quadratic action is the same as \eqref{Eq:SKAction}, however under the modified KMS transformation we find
	\begin{eqnarray}
		\label{Eq:SKActionTransModified}
		&\;& \int d^{3+1} x \;  \left[ - i (\mathcal{L} X)^{T} \bar{\Theta}^{T} \left( \begin{array}{c} J \\ V \end{array} \right)_{\mathrm{HS}}  + i ( \mathcal{L}  X)^{T} ( \bar{\Theta}^{T} \mathbbm{D}_{S} \bar{\Theta} - T \bar{\Theta}^{T} \mathbbm{N} \bar{\Theta}) ( \mathcal{L} X)  \right. \nonumber \\
		&\;& \left. \hphantom{ + \int d^{3+1} x \;  \left[ \right.} \vphantom{ \left( \begin{array}{c} J \\ V \end{array} \right)_{\mathrm{HS}}} + X_{a} ^{T} \left( - ( \bar{\Theta}^{T} \mathbbm{D} \bar{\Theta} + \mathbbm{D} ) + 2 T \bar{\Theta}^{T} \mathbbm{N} \bar{\Theta} \right) (\mathcal{L} X) \right. \nonumber \\
		&\;& \left. \hphantom{ + \int d^{3+1} x \;  \left[ \right.} \vphantom{ \left( \begin{array}{c} J \\ V \end{array} \right)_{\mathrm{HS}}} + X_{a}^{T} \left( \bar{\Theta}^{T} \left( \begin{array}{c} J \\ 0 \end{array} \right)_{\mathrm{HS}}  - \left( \begin{array}{c} J \\ 0 \end{array} \right)_{\mathrm{HS}}  \right) + i T X_{a}^{T} \left( \bar{\Theta}^{T} \mathbbm{N} \bar{\Theta} - \mathbbm{N} \right) X_{a} \right] \; . \qquad \; \; 
	\end{eqnarray}
Performing the same manipulations as in section \ref{sec:standardKMS} one finds that the operators $\mathbbm{D}$ and $\mathbbm{N}$ must satisfy modified versions of \eqref{Eq:OnsagerCasimir} i.e.
	\begin{subequations}
	\label{Eq:ModifiedOnsagerCasimir}
	\begin{eqnarray}
		\bar{\Theta}^{T} \mathbbm{D}[\Theta X] \bar{\Theta} &=& \mathbbm{D}^{\dagger}[X] \; , \\
		\bar{\Theta}^{T} \mathbbm{N}[\Theta X] \bar{\Theta} &=& \mathbbm{N}[X] \; ,
	\end{eqnarray}
	\end{subequations}
and the original fluctuation dissipation relationship \eqref{Eq:SimpleFDT}. If we can impose these relations, we can fit a Maxwell-Cattaneo style model into the Schwinger-Keldysh action. In passing, we note that once again one can add higher response terms such as 
	\begin{eqnarray}
		\mathcal{O}(X_{a}^3)
		&=& i \gamma \left( X_{a}^{T} M (X_{a} - i \mathcal{L}X) \right)^3 \, ,
		\; \; M[X] = \bar{\Theta}^{T} M[\Theta X] \bar{\Theta}  \, , \; \; M^{T} = M \, , \qquad
	\end{eqnarray}
to our action, whether they are KMS closed or not, following the strategy of section \ref{sec:standardKMS} and still find that positivity of entropy production is implied by positivity of $\mathbbm{N}$.}

{\ In the next two sections however we shall attempt to embed our hydrodynamics at orders one and two in derivatives respectively, into a Schwinger-Keldysh action with the transformation \eqref{Eq:ModifiedKMSX}.}

\subsubsection{Action at $\mathcal{O}(\partial)$ }

{\noindent Once we truncate the equations of motion \eqref{Eq:EquationsofMotion} to first order in derivatives, we immediately identify
	\begin{eqnarray}
	\label{Eq:O1modifieddecomp}
		\left( \begin{array}{c} J \\ V \end{array} \right)_{\mathrm{HS}} 
		&=& \left( \begin{array}{c} n u^{\mu} \\ 0 \end{array} \right) \; , \qquad
		 \mathbbm{D} = \left( \begin{array}{cc} (\sigma_{\mathrm{DC}} - r \tau) & \alpha \\ - \frac{\lambda r \tau}{\alpha} & \lambda ( 1 + \tau D )  \end{array} \right)\; , 
	\end{eqnarray}
where $\lambda \neq 0$ is a normalisation that drops out of the Maxwell-Cattaneo equation. Splitting the operator $\mathbbm{D}$ into self-adjoint and anti-self-adjoint pieces one finds
	\begin{subequations}
	\label{Eq:FirstOrderDSDA}
	\begin{eqnarray}
		\mathbbm{D}_{S} &=&  \left( \begin{array}{ccc} (\sigma_{\mathrm{DC}} - r \tau) & \; & \frac{1}{2} \left( \alpha - \frac{\lambda r \tau}{\alpha} \right) \\ \frac{1}{2} \left( \alpha - \frac{\lambda r \tau}{\alpha} \right)  & \; & \lambda - \frac{1}{2} D(\lambda \tau) \end{array} \right) \nonumber \\
		&=& \left( \begin{array}{ccc} (\sigma_{\mathrm{DC}} - r \tau) &\; &  \frac{1}{2} \left( \alpha - \frac{\lambda r \tau}{\alpha} \right) \\ \frac{1}{2} \left( \alpha - \frac{\lambda r \tau}{\alpha} \right)  &\; &  \lambda \end{array} \right) + \mathcal{O}(\partial) \; , \\
		\mathbbm{D}_{A} &=&   \left( \begin{array}{ccc} 0 & \; &  \frac{1}{2} \left( \alpha + \frac{\lambda r \tau}{\alpha} \right) \\ - \frac{1}{2} \left( \alpha + \frac{\lambda r \tau}{\alpha} \right)  &  \; & \lambda \tau D + \frac{1}{2} D(\lambda \tau) \end{array} \right)  \nonumber \\
		&=& \left( \begin{array}{ccc} 0 & \; & \frac{1}{2} \left( \alpha + \frac{\lambda r \tau}{\alpha} \right) \\ - \frac{1}{2} \left( \alpha + \frac{\lambda r \tau}{\alpha} \right)  & \; & \lambda \tau D \end{array} \right)  + \mathcal{O}(\partial)  \; . 
	\end{eqnarray}
	\end{subequations}
The term $D(\lambda \tau) \sim \mathcal{O}(\partial)$ is not relevant at this order in derivatives, but will reappear at next order. We see immediately that $\mathbbm{D}_{S}$ matches the expression that occurs in our entropy current argument of section \ref{sec:orderoneentropy} if we identify $\lambda = \rho_{v}/\tau$. Moreover, $\mathbbm{D}_{S}$ is equivalent to $T \mathbbm{N}$ by construction. As our action is quadratic this further means that $\mathbbm{N}>0$ as an operator if the action is to be unitary.}

{\ It should be noted that in writing \eqref{Eq:FirstOrderDSDA} we have not yet imposed the modified Onsager-Casimir relations \eqref{Eq:ModifiedOnsagerCasimir}. We quickly find that \eqref{Eq:ModifiedOnsagerCasimir} constrains the off-diagonal pieces of $\mathbbm{D}_{S}$. This conclusion is in apparent tension with the explicit computation of the retarded Green's functions in section \ref{sec:orderoneentropy} as it requires
\begin{eqnarray}
	\label{Eq:Constraint}
   	 \rho_{v} &=&  \frac{\alpha^{2}}{r}\; ,
\end{eqnarray}
which we did not previously need to impose. Equation \eqref{Eq:Constraint} can be understood as a constraint to allow the Maxwell-Cattaneo equations to be embedded in a Schwinger-Keldysh theory with the modified KMS symmetry of \cite{Jain:2023obu} - in particular we have assigned a particular time reversal behaviour to the auxiliary field. The observable conductivities \eqref{Eq:CompactConductivities} did not exhibit this same obstruction because the auxiliary field is eliminated when computing response. In particular, the real current response is given by the Schur complement
\begin{eqnarray}
    \mathbbm{D}_{\mathrm{eff}}
    =
    \mathbbm{D}_{JJ}
    -
    \mathbbm{D}_{Jv}
    \mathbbm{D}_{vv}^{-1}
    \mathbbm{D}_{vJ},
\end{eqnarray}
rather than by the individual blocks of $\mathbbm{D}$ themselves. Reciprocity of the physical Green's functions consequently constrains only this reduced operator.}

{\ Since the auxiliary field does not itself represent an observable hydrodynamic current, its transformation properties need not in principle be fixed by the reciprocity relations obeyed by the conserved current\footnote{A closely related phenomenon occurs already in relaxed hydrodynamics, where correlators are generically not time-reversal covariant even when Onsager-reciprocal, and covariance selects a minimal framework \cite{Amoretti:2023hbc}.}. It is our choice of embedding that is leading to such a constraint, and whether there are other embeddings that alleviate this restriction is an open question for future work. Nevertheless, it should be noted that there is enough room in our constitutive relations even when \eqref{Eq:Constraint} is imposed to capture the leading holomorphic terms necessary to match the retarded Green's functions \cite{Amoretti:2025kem}. At the orders we consider, the differences manifest in the non-linear terms of the constitutive relations.}

\subsubsection{Action at $\mathcal{O}(\partial^2)$ }\label{sec:secondorderSK}

{\noindent At second order in derivatives we encounter a new and interesting phenomenon for the inverse problem: the non-linear terms which mix the auxiliary field and physical current sectors cannot be arbitrarily assigned to $\mathbbm{D}_{S}$ or $\mathbbm{D}_{A}$. Before discussing this, let us first isolate the corrections to the hydrostatic expressions which take the form
	\begin{eqnarray}
			\left. \left( \begin{array}{c} J \\ V \end{array} \right)_{\mathrm{HS}} \right|_{\mathcal{O}(\partial^2)} 
		&=& \left( u^{\mu} \left[ \chi_{E} \nabla_{\nu} E^{\nu} + \frac{\partial \chi_{E}}{\partial \mu} E^{\nu} \nabla_{\nu}^{\perp} \mu \right] - \chi_{E} DE^{\mu} - \frac{\partial \chi_{E}}{\partial \mu} D \mu E^{\mu} \right. \nonumber \\
		&\;& \left. \hphantom{ \left( \right.} + \chi_{B} \epsilon^{\mu \nu \rho \sigma} u_{\rho} \nabla_{\nu} \mathcal{B}_{\sigma} +  \frac{\partial \chi_{B}}{\partial \mu} \epsilon^{\mu \nu \rho \sigma} u_{\rho} ( \nabla_{\nu}^{\perp} \mu ) \mathcal{B}_{\sigma} \right) \left( \begin{array}{c} 1 \\ 0 \end{array} \right) \; .
	\end{eqnarray}
This expression is standard from charged hydrodynamics in an external electromagnetic field \cite{Kovtun:2016lfw}.}

{\ Turning now to $\mathbbm{D}$ we note that there are now two types of contribution for the constitutive relations in \eqref{Eq:LinEquationsofMotion}: terms that are linear in $\mathcal{L}X$ which have an unambiguous assignment to the self-adjoint or anti-self-adjoint part of $\mathbbm{D}$, and terms non-linear in $\mathcal{L}X$ which have ambiguous assignment. To parameterise this latter ambiguity, let us introduce independent parameters $\varepsilon_{1}$ and $\varepsilon_{2}$ so that we subsequently find the following family of operators:
\begin{subequations}
\label{Eq:OrderTwoEpsilonD}
\begin{eqnarray}
\left.\mathbbm{D}\right|_{\mathcal{O}(\partial^{2})}
&=&
\left(\begin{array}{ccc}
\mathbbm{A}_{\boldsymbol{\varepsilon}} & \; &
\mathbbm{B}_{\boldsymbol{\varepsilon}}\\
0& \; & 0
\end{array}\right)\; ,\\
(\mathbbm{A}_{\boldsymbol{\varepsilon}})^{\mu}{}_{\nu}
&=&\mathfrak{c}_{1} u^{\mu} \nabla_{\nu}^{\perp} + \mathfrak{c}_{2}\,\Pi^{\mu}{}_{\nu}D
+u^{\mu}\left[
 \mathfrak{b}_{5}E_{\nu}+\mathfrak{b}_{6} (E_{\nu}-\partial^{\perp}_{\nu}\mu)
 +(\varepsilon_{1}\mathfrak{b}_{2}+\varepsilon_{2}\mathfrak{b}_{0}) v_{\nu}
\right] \nonumber \\
&\;& +\mathfrak{c}_{3} \epsilon^{\mu\rho}{}_{\nu\sigma}u_{\rho}B^{\sigma} \; ,\\
(\mathbbm{B}_{\boldsymbol{\varepsilon}})^{\mu}{}_{\nu}
&=&  \mathfrak{b}_{1}u^{\mu}\partial^{\perp}_{\nu}
+u^{\mu}\left[
 \mathfrak{b}_{2}(E_{\nu}-\varepsilon_{1} (E_{\nu}-\partial^{\perp}_{\nu}\mu))
 +(1-\varepsilon_{2})\mathfrak{b}_{0} (E_{\nu}-\partial^{\perp}_{\nu}\mu)
\right] \nonumber\\
&& +\mathfrak{b}_{3}  u^{\mu} v_{\nu}
+\mathfrak{b}_{4} \epsilon^{\mu\rho}{}_{\nu\sigma}u_{\rho}B^{\sigma}; .
\end{eqnarray}
\end{subequations}
Crucially however, when these operators act on $\mathcal{L}X=(E - \partial_{\perp} \mu,v)^{T}$  we find
\begin{eqnarray}
&\;& (\mathbbm{A}_{\boldsymbol{\varepsilon}})\indices{^\mu_\nu} (E^{\nu}-\partial_{\perp}^{\nu}\mu)+
(\mathbbm{B}_{\boldsymbol{\varepsilon}})\indices{^\mu_\nu} v^{\nu} \nonumber \\
&=& u^{\mu} \left[\mathfrak{b}_{5}E_{\nu}  (E^{\nu} - \partial_{\perp}^{\nu} \mu) +\mathfrak{b}_{2} E_{\nu} v^{\nu}
+\mathfrak{b}_{6}  (E - \partial_{\perp} \mu) ^{2}+\mathfrak{b}_{0}  (E^{\nu} - \partial_{\perp}^{\nu} \mu)  v_{\nu} +\mathfrak{b}_{3} v^{2}\right] \nonumber \\
&\;& + \mathfrak{c}_{2}\,\Pi^{\mu}{}_{\nu} D (E^{\nu} - \partial_{\perp}^{\nu} \mu) +  \mathfrak{b}_{1}u^{\mu}\partial^{\perp}_{\nu} v^{\nu} +\mathfrak{c}_{3} \epsilon^{\mu\rho}{}_{\nu\sigma}u_{\rho}B^{\sigma} (E^{\nu} - \nabla^{\nu}_{\perp} \mu) \nonumber \\
&\;& +\mathfrak{b}_{4} \epsilon^{\mu\rho}{}_{\nu\sigma}u_{\rho}B^{\sigma} v_{\mu} \; , \qquad
\end{eqnarray}
which is independent of $\varepsilon_{1}$ and $\varepsilon_{2}$ and thus the generic constitutive relations are unchanged before embedding in Schwinger-Keldysh.}

{\ For arbitrary $\varepsilon_{1}$ and $\varepsilon_{2}$, the corresponding self-adjoint and anti-self-adjoint corrections are
\begin{subequations}
\label{Eq:OrderTwoEpsilonDSDA}
\begin{eqnarray}
\left.\mathbbm{D}_{S}\right|_{\mathcal{O}(\partial^{2})}
&=&\frac{1}{2}
\left(\begin{array}{ccc}
\mathbbm{A}_{\boldsymbol{\varepsilon}}+
\mathbbm{A}_{\boldsymbol{\varepsilon}}^{\dagger} & \;
&\mathbbm{B}_{\boldsymbol{\varepsilon}}\\
\mathbbm{B}_{\boldsymbol{\varepsilon}}^{\dagger}& \; & - D(\lambda \tau)
\end{array}\right)\; ,\\
\left.\mathbbm{D}_{A}\right|_{\mathcal{O}(\partial^{2})}
&=&\frac{1}{2}
\left(\begin{array}{ccc}
\mathbbm{A}_{\boldsymbol{\varepsilon}}-
\mathbbm{A}_{\boldsymbol{\varepsilon}}^{\dagger} & \;
&\mathbbm{B}_{\boldsymbol{\varepsilon}}\\
-\mathbbm{B}_{\boldsymbol{\varepsilon}}^{\dagger}& \; & D(\lambda \tau)
\end{array}\right)\; .
\end{eqnarray}
\end{subequations}
When embedded in our Schwinger-Keldysh formulation, these operators must satisfy the relevant Onsager-Casimir constraints \eqref{Eq:ModifiedOnsagerCasimir}. It is natural to ask whether the ambiguity encoded by the parameters $\varepsilon_{1,2}$ is removed by these relations. Applying the relations directly to the enlarged operator $\mathbbm{D}$ leads to an action on the individual blocks $\mathbbm{A}_{\boldsymbol{\varepsilon}}$ and $\mathbbm{B}_{\boldsymbol{\varepsilon}}$ and we discover that we must set
	\begin{eqnarray}
		\mathbbm{B}_{\boldsymbol{\varepsilon}} 
		&=&  \mathfrak{b}_{1}u^{\mu}\partial^{\perp}_{\nu}
+u^{\mu}\left[
 \mathfrak{b}_{2}(E_{\nu}-\varepsilon_{1} (E_{\nu}-\partial^{\perp}_{\nu}\mu))
 +(1-\varepsilon_{2})\mathfrak{b}_{0} (E_{\nu}-\partial^{\perp}_{\nu}\mu)
\right] \nonumber\\
&& + \mathfrak{b}_{3} u^{\mu} v_{\nu}
+\mathfrak{b}_{4} \epsilon^{\mu\rho}{}_{\nu\sigma}u_{\rho}B^{\sigma}  = 0 \; . 
	\end{eqnarray}
For arbitrary field configurations one must then choose
	\begin{eqnarray}
		\mathfrak{b}_{1} = \mathfrak{b}_{2} = \mathfrak{b}_{3} = \mathfrak{b}_{4} = 0 \; .
	\end{eqnarray}
Vanishing of $\mathbbm{B}_{\boldsymbol{\varepsilon}}$ for arbitrary field configurations further imposes $(1-\varepsilon_{2})\mathfrak{b}_{0}=0$, while the residual term $\varepsilon_{2}\mathfrak{b}_{0} v_{\nu}$ in $\mathbbm{A}_{\boldsymbol{\varepsilon}}$ is removed by the constraints on that block; together these force $\mathfrak{b}_{0}=0$. As for $\mathbbm{A}_{\boldsymbol{\varepsilon}}$, we find that the modified Onsager-Casimir constraints impose that
	\begin{eqnarray}
		\mathfrak{b}_{5}=\mathfrak{b}_{6}=0 \; ,
	\end{eqnarray}
and thus all $\varepsilon_{1,2}$ ambiguity has dropped from our expressions which reduce to
\begin{eqnarray}
\label{Eq:ReducedAB}
(\mathbbm{A}_{\boldsymbol{\varepsilon}})^{\mu}{}_{\nu}
&=& \mathfrak{c}_{1} u^{\mu} \nabla_{\nu}^{\perp} + \mathfrak{c}_{2}\,\Pi^{\mu}{}_{\nu}D
+\mathfrak{c}_{3} \epsilon^{\mu\rho}{}_{\nu\sigma}u_{\rho}B^{\sigma} \; , \qquad 
(\mathbbm{B}_{\boldsymbol{\varepsilon}})^{\mu}{}_{\nu}
= 0 \;  .
\end{eqnarray}
}

{\ We see that in writing \eqref{Eq:ReducedAB}, all couplings $\mathfrak{b}_{i}$ between the auxiliary field and the physical current have been set to zero at this order in derivatives. We expect this radical reduction in transport coefficients to be a feature of the order-two analysis, rather than generic. At order three, for example, the lower-left quadrant of the $\mathbbm{D}$ matrix receives corrections, and it is precisely the vanishing of this quadrant at second order on which the cancellation above depends. Moreover, the following operator is invariant under modified KMS, possesses the $\varepsilon$ ambiguity, is third order in derivatives and vanishes when evaluated on the thermal forces that make up the constitutive relations
	\begin{eqnarray}
		\label{Eq:OrderThreeAmbiguity}
		\varepsilon \left( \begin{array}{ccc} 
			v^{\mu} v_{\nu} & \; & - v \cdot (E - \partial_{\perp} \mu) \delta\indices{^\mu_\nu} \\
			 - v \cdot (E - \partial_{\perp} \mu) \delta\indices{^\mu_\nu}  & \; & ( E^{\mu} -\partial_{\mu}^{\perp} \mu ) ( E_{\nu} -\partial_{\nu}^{\perp} \mu )
		\end{array} \right) \; . 
	\end{eqnarray}
It is a true ambiguity of the inverse problem and thus the existence of a space of Gaussian theories with equivalent non-linear hydrodynamics, but distinct noise correlators is a non-trivial issue worthy of future development. This ambiguity is conceptually related to the observation of \cite{Jain:2023obu} that the auxiliary ultraviolet sector can admit inequivalent realisations of dynamical KMS symmetry. The present ambiguity is sharper: within a fixed modified-KMS realisation, the deformation \eqref{Eq:OrderThreeAmbiguity} leaves the nonlinear constitutive relations unchanged while modifying the Gaussian noise kernel. This specific inverse-problem ambiguity was not identified in \cite{Jain:2023obu}. 

\subsubsection{Integrating out the auxiliary vector field }
\label{Sec:IntegratingOutAdditionalField}

{\noindent The modified affine KMS symmetry acts on the enlarged field space containing both the ordinary hydrodynamic field and the additional relaxation field. We now show that, regardless of imposing a definitive time transformation eigenvalue on $\zeta$, after integrating out the auxiliary field the resulting effective action for the observable hydrodynamic sector will still be invariant under the standard affine KMS transformation. Moreover, we shall show how integrating out the auxiliary response field affects the resultant theory. Coloured noise induced by a Cattaneo-type relaxation time was first observed in
 \cite{Kapusta:2014dja} by imposing the fluctuation-dissipation theorem directly on the
 linearised theory; here the same structure follows from the modified KMS symmetry of the
 enlarged action.}

{\ For this purpose, consider \eqref{Eq:SKAction} and let us decompose the matrix $\mathbbm{D}$ into block form
\begin{eqnarray}
\label{Eq:DBlockForm}
\mathbbm{D}
&=&
\left(
\begin{array}{ccc}
\mathbbm{D}_{JJ} & \; & \mathbbm{D}_{Jv} \\
\mathbbm{D}_{vJ} &\; & \mathbbm{D}_{vv}
\end{array}
\right) \; . 
\end{eqnarray}
As the action that we arrived at in section \ref{sec:secondorderSK} is at most quadratic in the auxiliary field $\zeta$, integration over it imposes the constraint
\begin{eqnarray}
\mathbbm{D}_{Jv}^{\dagger}B_{a}
+\mathbbm{D}_{vv}^{\dagger}\zeta_{a}
&=&0 .
\label{Eq:ZetaConstraint}
\end{eqnarray}
Provided that $\mathbbm{D}_{vv}$ is invertible on the space of fluctuations under consideration, this constraint can be solved for the auxiliary response field
\begin{eqnarray}
\zeta_{a}
&=&
-\left(\mathbbm{D}_{vv}^{\dagger}\right)^{-1}
\mathbbm{D}_{Jv}^{\dagger}B_{a} .
\label{Eq:ZetaASolution}
\end{eqnarray}
Substitution into the real part of the action then produces the reduced response operator
\begin{eqnarray}
\mathbbm{D}_{\mathrm{eff}}
&=&
\mathbbm{D}_{JJ}
-\mathbbm{D}_{Jv}\mathbbm{D}_{vv}^{-1}\mathbbm{D}_{vJ} .
\label{Eq:DEffective}
\end{eqnarray}
As expected, this is the Schur complement of the relaxation block $\mathbbm{D}_{vv}$.}

{\ To evaluate the reduced noise kernel, define
\begin{eqnarray}
\mathbbm{L}_{a}
&=&
\left(
\begin{array}{c}
\mathbbm{1} \\
-\left(\mathbbm{D}_{vv}^{\dagger}\right)^{-1}
\mathbbm{D}_{Jv}^{\dagger}
\end{array}
\right) ,
\qquad
X_{a}=\mathbbm{L}_{a} B_{a} .
\label{Eq:LaDefinition}
\end{eqnarray}
The effective noise operator is consequently
\begin{eqnarray}
\mathbbm{N}_{\mathrm{eff}}
&=& \frac{1}{T}
\mathbbm{L}_{a}^{\dagger}
\mathbbm{D}_{S}
\mathbbm{L}_{a} 
= \frac{1}{2 T}
\left[
\mathbbm{D}_{JJ}
-\mathbbm{D}_{Jv}\mathbbm{D}_{vv}^{-1}\mathbbm{D}_{vJ}
+\mathbbm{D}_{JJ}^{\dagger}
-\mathbbm{D}_{vJ}^{\dagger}
\left(\mathbbm{D}_{vv}^{\dagger}\right)^{-1}
\mathbbm{D}_{Jv}^{\dagger}
\right]
\nonumber \\
&=&
\frac{1}{2 T}
\left(
\mathbbm{D}_{\mathrm{eff}}
+\mathbbm{D}_{\mathrm{eff}}^{\dagger}
\right) .
\end{eqnarray}
Consequently, the reduced action takes the form
\begin{eqnarray}
S
&=&
\int d^{3+1}x \;
\left[ B_{a}^{T} \left( J_{\mathrm{HS}} +  \mathbbm{D}_{\mathrm{eff}}
\mathcal{L}_{\beta}B \right)  +\frac{i}{2}B_{a}^{T}
\left(
\mathbbm{D}_{\mathrm{eff}}
+\mathbbm{D}_{\mathrm{eff}}^{\dagger}
\right)
B_{a}
\right] .
\label{Eq:ReducedStandardKMSAction}
\end{eqnarray}
This is precisely the quadratic Schwinger--Keldysh action satisfying the standard fluctuation--dissipation relation. It is therefore invariant under the standard affine KMS transformation
\begin{eqnarray}
B_{a}
&\longrightarrow&
\Theta[ B_{a}-i\mathcal{L}_{\beta}B ] ,
\label{Eq:ReducedStandardKMS}
\end{eqnarray}
together with the appropriate action of the discrete transformation on $B$ and on the spacetime arguments. This action \eqref{Eq:ReducedStandardKMSAction} is the effective non-local action discussed in the introduction.}

{\ We can now see that even in the model where we have integrated out the auxiliary field $\zeta$ and its noise, we can still determine their presence. In the linearised deterministic correlators the auxiliary field manifests as a gapped pole (see the \eqref{Eq:Nonlocalconstitutive} and section \ref{sec:linearised}). On the other hand, the corresponding response field $\zeta_{a}$ has introduced a non-local pole into the effective noise correlator $\mathbbm{N}_{\mathrm{eff}}$. While higher derivative hydrodynamic terms generically convert white noise at order zero in derivatives to coloured noise, such corrections remain local. Integrating out $\zeta_{a}$ however has added non-local memory effects. This is analogous to the familiar influence-functional mechanism whereby integrating out environmental degrees of freedom generates non-local dissipation and noise kernels \cite{Kaplanek:2025moq}.}

\subsection{Stochastic quasi-hydrodynamics of the embedded Maxwell-Cattaneo model}\label{sec:stochastic}

{\noindent Finally, we obtain the nonlinear stochastic Maxwell--Cattaneo equations associated with the Gaussian Schwinger--Keldysh embedding constructed above. We first explain how the modified dynamical KMS transformation constrains the probability of joint noise histories. We then eliminate the Maxwell--Cattaneo field at the level of the stochastic equations and show explicitly that its associated noise contributes to observable charge fluctuations. Throughout this section we restrict to actions quadratic in the response fields and hence to Gaussian noise in the enlarged local theory. Higher response-field interactions can describe non-Gaussian noise, but a finite polynomial deformation never corresponds to a positive stochastic probability, by the classic theorem of Marcinkiewicz \cite{Marcinkiewicz:1939wy,Pawula:1967zz}. The stronger positivity conditions required for a genuine stochastic interpretation of non-Gaussian Schwinger-Keldysh noise, together with explicit positive completions of truncated noise data, are studied in \cite{Amoretti:2026positivenoise}.}

{\ We seek to introduce stochastic noise fields $\eta$ which are defined to satisfy the following relationship
	\begin{eqnarray}
		Z[X_{a};X] = \exp\left[i \int d^{3+1} x \;  X_{a}^{T} \left( \begin{array}{c} J \\ V \end{array} \right)\right] \int D\eta \; P[\eta|X] e^{\sqrt{2 T} i \int d^{3+1}x \; X_{a}^{T} \eta} \; ,
	\end{eqnarray}
with $P[\eta|X] \geq 0$ a positive, normalised functional at fixed $X$ representing the probability of that configuration of the $\eta$ variables given the value of $X$. For our Gaussian actions \eqref{Eq:SKAction}, we can linearise the quadratic behaviour of the response fields $B_{a}$ and $\zeta_{a}$ through stochastic variables $\eta$ by performing a Hubbard-Stratonovich transformation. In particular, we employ the identity
	\begin{eqnarray}
		&\;& e^{- T \int d^{3+1}x \; X_{a}^{T} \mathbbm{N} X_{a} } \nonumber \\
		 &=& \mathcal{N}[X] \int D \eta e^{- \frac{1}{2} \int d^{3+1}x d^{3+1} y\; \left( \eta^{T}(x) \mathbbm{N}^{-1}(x,y) \eta(y) \right) + i \sqrt{2T}  \int d^{3+1}x  \left( X_{a}^{T} \eta  \right) } \; , \qquad
	\end{eqnarray}
where we have assumed that $\mathbbm{N}$ is positive, self-adjoint and invertible, $(\eta^{T})^{\mu} = ( \eta_{B}, \eta_{\zeta} )^{\mu}$ and $\mathcal{N}[X]>0$ is a non-trivial normalisation dependent on the fields. The resultant Schwinger-Keldysh generating functional $Z$ looks like
	\begin{eqnarray}
		&\;& Z = \int D X_{a} D X \; e^{i S} = \int D \eta DX\;  P[\eta|X] \delta\left( \left( \begin{array}{c} \partial \cdot J  \\ V \end{array} \right) + \sqrt{2 T}  \left( \begin{array}{c} \partial \cdot \eta_{B} \\ \eta_{\zeta} \end{array} \right)\right) \; , \nonumber \\
		 &\;& P[\eta|X] = \mathcal{N}[X] e^{\left( - \frac{1}{2} \int d^{3+1}x d^{3+1}y \; \eta^{T}(x) \mathbbm{N}^{-1}(x,y) \eta(y) \right)} \; , \qquad \int D \eta \; P[\eta|X] = 1 \; ,
	\end{eqnarray}
where we have set $B_{a\mu} = \partial_{\mu} \varphi_{a}$. Our equations of motion \eqref{Eq:EquationsofMotion} are then supplemented by stochastic source terms
\begin{subequations}
\label{Eq:GenericStochasticEqns}
\begin{eqnarray}
\nabla_{\mu} J^{\mu} &=& - \sqrt{2T}\, \nabla_{\mu} \eta_{B}^{\mu} \; ,\\
v^{\mu}+\tau Dv^{\mu}-\frac{r\tau}{\alpha}
\left(E^{\mu}-\partial_{\perp}^{\mu}\mu\right)
&=& - \frac{\sqrt{2T}\, r \tau}{\alpha^2}\,\eta_{\zeta}^{\mu}
+\mathcal{O}(\partial^{3})\; ,
\end{eqnarray}
where the overall normalisation of the Maxwell-Cattaneo equation noise $\eta_{\zeta}^{\mu}$ follows from $\lambda$ introduced in \eqref{Eq:O1modifieddecomp} and the noise fields $\eta$ are drawn from a distribution satisfying
	\begin{eqnarray}
		\label{Eq:ThermalNoiseDistribution}
		\langle \eta_{I}(x) \eta_{J}(y) \rangle_{X} &=& \mathbbm{N}_{IJ}[X](x,y) \; .
	\end{eqnarray}
\end{subequations}
To solve equations \eqref{Eq:GenericStochasticEqns} a discretization prescription is required \cite{dePirey:2022vyg}.}

{\ Having now introduced the stochastic measure $P[\eta|X]$, we can give an interpretation to the modified KMS transformation of \eqref{Eq:ModifiedKMSX} and compare it to the standard case \eqref{Eq:KMStrans}. Assuming that the action is KMS invariant we have a relationship between the original and time reversed generating functional
	\begin{eqnarray}
			Z_{\Theta}[\Theta(X_{a} - i \mathcal{L}_{\beta}X);\Theta X]	
		&=& Z[X_{a} ;X] \; .
	\end{eqnarray}
Assuming functional invariance of the measure under $\eta = \Theta \eta'$ we arrive at
	\begin{eqnarray}
		&\;& Z_{\Theta}[\Theta(X_{a} - i \mathcal{L}_{\beta}X);\Theta X] \nonumber \\
		&=& \exp\left[i \int d^{3+1} x \; \left( X_{a}^{T} \Theta^{T} \left( \begin{array}{c} J \\ V \end{array} \right) - i (\mathcal{L}_{\beta} X) \Theta^{T} \left( \begin{array}{c} J \\ V \end{array} \right) \right) \right] \times \nonumber \\ 
		&\;& \int D\eta' \; P_{\Theta}[\eta'|\Theta X] e^{ \sqrt{2T} i \int d^{3+1}x \;  (X_{a} - i \mathcal{L}_{\beta}X))^{T} \eta'} \nonumber \\
		&=& \exp\left[i \int d^{3+1} x \; \left( X_{a}^{T} \left( \begin{array}{c} J \\ V \end{array} \right) - i (\mathcal{L}_{\beta} X) \left( \begin{array}{c} J \\ V \end{array} \right) \right) \right] \times \nonumber \\
		&\;&  \int D\eta \; \left( P_{\Theta}[\eta|\Theta X] e^{ \sqrt{2T}   \int d^{3+1}x \;  (\mathcal{L}_{\beta}X)^{T} \eta}  \right) e^{\sqrt{2T} i \int d^{3+1}x \;  X_{a}^{T} \eta} \; .
	\end{eqnarray}
Equality of the Fourier transforms implies the corresponding relation between the
conditional noise measures,
\begin{equation}
 \frac{P[\eta\,|\,X]}
      {P_\Theta[\Theta\eta\,|\,\Theta X]}
 =
 \exp\!\left[
   \int d^{3+1}x\,(\mathcal L_\beta X)^T \left(  \left( \begin{array}{c} J \\ V \end{array} \right) + \sqrt{2 T} \eta \right)
 \right] \; .
\label{eq:standard-noise-ratio}
\end{equation}
The normalization of the two measures consequently implies the associated integral fluctuation relation. For the modified transformation, the same argument holds with
\begin{equation}
 \mathcal L X=
 \begin{pmatrix}
   \mathcal L_\beta B\\[2pt]
   \zeta
 \end{pmatrix},
\end{equation}
and hence
\begin{equation}
 \frac{P[\eta\,|\,X]}
      {P_\Theta[\Theta\eta\,|\,\Theta X]}
 =
 \exp\!\left[
   \int d^{3+1}x\,(\mathcal L X)^T \left(  \left( \begin{array}{c} J \\ V \end{array} \right) + \sqrt{2 T} \eta \right)
 \right].
\label{eq:modified-noise-ratio}
\end{equation}
Thus the modified KMS transformation constrains the relative weights of forward and reversed joint noise histories with the term $\zeta_{\mu}$ as opposed to $(\mathcal{L}_{\beta} \zeta)_{\mu}$.}

{\ While \eqref{eq:modified-noise-ratio} is convenient for its relative simplicity, the more common object to examine is the ratio of hydrodynamic histories which defines local detailed balance. If we wanted to convert \eqref{eq:modified-noise-ratio} into a statement about hydrodynamic histories we would first need to solve
\begin{eqnarray}
	\label{eq:Noisetohistory}
	   \left( \begin{array}{c} \nabla \cdot \eta_{B} \\ \eta_{\zeta} \end{array} \right) = - \frac{1}{ \sqrt{2 T}} \left( \begin{array}{c} \nabla \cdot J  \\ V \end{array} \right)
\end{eqnarray}
which means determining the behaviour of the corresponding noise field given a particular hydrodynamic history. If this map were one to one we could then write
\begin{equation}
 P[X]
 =
 P\left[\left( \begin{array}{c} \partial \cdot J  \\ V \end{array} \right) | X \right] \,\mathcal J[X],
\end{equation}
where we have substituted into $P[\eta|X]$ using \eqref{eq:Noisetohistory} and $\mathcal J[X]$ is the noise-to-history Jacobian. In such a situation one would then find
\begin{equation}
 \frac{P[X]}{P_\Theta[\Theta X]}
 =
 \frac{P[\mathcal E[X]\,|\,X]}
      {P_\Theta[\mathcal E[\Theta X]\,|\,\Theta X]}
 \frac{\mathcal J[X]}{\mathcal J_\Theta[\Theta X]} .
\label{eq:history-ratio}
\end{equation}
relating noise probabilities to detailed balance. However, the equation \eqref{eq:Noisetohistory} presents difficulties; for example, for field-dependent noise the Jacobian depends on the stochastic prescription. Only after showing that the regulated forward and reversed Jacobians agree may the probability ratio of hydrodynamic histories be identified directly with the noise-history ratio. Thus we find the interpretation in terms of noise ratios \eqref{eq:modified-noise-ratio} more workable in this context.}

{\ Having discussed interpretation, we now remind ourselves that the component $\eta_\zeta^\mu$ is auxiliary, but this does not imply that it is physically irrelevant. This can be seen directly by eliminating $v^\mu$ from the stochastic equations \eqref{Eq:GenericStochasticEqns}. Once more using $\mathbbm{K}$ defined in \eqref{Eq:RelaxationOperator}, we see that
\begin{equation}
 v^\mu
 =
 (\mathbbm{K}^{-1})^\mu{}_\nu
 \left[
   \frac{r\tau}{\alpha}
      \bigl(E^\nu-\partial_\perp^\nu\mu\bigr)
   -\sqrt{2T}\,\frac{r\tau}{\alpha^2}\eta_\zeta^\nu
 \right].
\label{eq:v-stochastic-solution}
\end{equation}
Since the physical current contains the term $\alpha v^\mu$, substitution into the
continuity equation gives a reduced stochastic equation of the form
\begin{equation}
 \nabla_\mu J_{\rm red}^\mu[B]
 =
 -\sqrt{2T}\,\nabla_\mu\eta_{\rm eff}^\mu ,
 \qquad
 \eta_{\rm eff}^\mu
 =
 \eta_B^\mu
 -\frac{r\tau}{\alpha}
   (\mathbbm{K}^{-1})^\mu{}_\nu\eta_\zeta^\nu .
\label{eq:effective-noise}
\end{equation}
Here $J_{\rm red}^\mu$ contains the corresponding deterministic behaviour.
Equation~\eqref{eq:effective-noise} shows that $\eta_\zeta^\mu$ cannot in general
be absorbed into a local redefinition of $\eta_B^\mu$: it is ``filtered'' by the inverse
relaxation operator. In Fourier space, for the linearised operator
\begin{equation}
 \mathbbm{K}^{-1}(\omega,\vec{k})
 \sim \frac{1}{1-i\omega\tau+\ldots},
\end{equation}
so the effective covariance contains the Maxwell-Cattaneo relaxation pole. Thus leading order white noise in the localised description, which becomes coloured at higher derivatives, becomes coloured noise with memory in the reduced description.}

{\ Having identified the thermal noise distribution \eqref{Eq:ThermalNoiseDistribution} and demonstrated that both stochastic variables represent physical noise, we can now summarise the stochastic hydrodynamic theory compatible with our Gaussian Schwinger-Keldysh actions, the existence of holomorphic terms and the modified KMS transformation of \cite{Jain:2023obu}. The non-linear equations of motion are
	\begin{subequations}
	\label{Eq:StochasticEquationsofMotion}
	\begin{eqnarray}
		\label{Eq:StochasticCurrent}
			v^{\mu}  + \tau D v^{\mu} + \frac{r \tau}{\alpha} \partial^{\mu}_{\perp} \mu  
		&=& \frac{r \tau}{\alpha} E^{\mu} - \frac{\sqrt{2T}  r \tau}{\alpha^2}\,\eta_{\zeta}^{\mu} +  \mathcal{O}(\partial^3) \; , \qquad \\
		\label{Eq:StochasticCharge}
		   	\chi_{\rho \rho} D \mu + \alpha \partial^{\perp}_{\mu} v^{\mu}  + \frac{\partial \alpha}{\partial \mu} v^{\mu} \partial_{\mu}^{\perp} \mu &-& (\sigma_{\mathrm{DC}} - r \tau)  \partial^{2}_{\perp} \mu  - \frac{\partial (\sigma_{\mathrm{DC}} - r \tau)}{\partial \mu} (\partial_{\perp} \mu)^2
			    \nonumber \\
			   &=& - (\sigma_{\mathrm{DC}} - r \tau) \partial_{\mu}^{\perp} E^{\mu} - \frac{\partial (\sigma_{\mathrm{DC}} - r \tau)}{\partial \mu} E^{2}   \nonumber \\
			   &\;& + \frac{\partial (\sigma_{\mathrm{DC}} - r \tau)}{\partial \mu} E^{\mu} ( E_{\mu} - \partial_{\mu}^{\perp} \mu )   - \sqrt{2T} \partial_{\mu} \eta_{B}^{\mu}  \nonumber \\
			   &\;& + \mathcal{O}(\partial^3) \; . \qquad 
	\end{eqnarray}
where the $U(1)$ charge current is given by 
	\begin{eqnarray}
		\label{Eq:StochasticChargeCurrentConstitutive}
		J^{\mu} &=& \left[ n + \chi_{E} \partial_{\nu}^{\perp} E^{\nu} + \mathfrak{c}_{1} \left( \partial_{\nu}^{\perp} E^{\nu} - \partial_{\perp}^2 \mu \right) 
		 +  \frac{\partial \chi_{E}}{\partial \mu} E^2 + \frac{\partial \chi_{E}}{\partial \mu} E^{\nu} ( E_{\nu} - \partial_{\nu}^{\perp} \mu )    \right. \nonumber \\
	&\;& \left.  + \frac{1}{2} \frac{\partial \chi_{B}}{\partial \mu} \mathcal{B}^{2} \right] u^{\mu} 
		+ (\sigma_{\mathrm{DC}} - r \tau) (E^{\mu} - \partial_{\perp}^{\mu} \mu) + \alpha v^{\mu}  
		- ( \chi_{E} - \mathfrak{c}_{2} ) D E^{\mu}
		  \nonumber \\
	&\;& + \epsilon^{\mu \nu \rho \sigma} u_{\nu} \left( \chi_{B} \partial^{\perp}_{\rho} + \frac{\partial \chi_{B}}{\partial \mu}  E_{\rho} 
		 + \left( \mathfrak{c}_{3}  - \frac{\partial \chi_{B}}{\partial \mu}  \right) ( E_{\rho} - \partial_{\rho}^{\perp} \mu)   \right) \mathcal{B}_{\sigma} \nonumber \\
	&\;& + \mathcal{O}(\partial^3) \; . 
	\end{eqnarray}
with the constraints
\begin{eqnarray}
\sigma_{\mathrm{DC}}\geq r \tau,
\qquad
\tau>0,
\qquad
r>0 \; .
\end{eqnarray}
The corresponding entropy current has the form
	\begin{eqnarray}
S^\mu
&=&
S_{\mathrm{can.}}^\mu
+S_{\mathrm{HS}}^\mu
+\frac{\alpha^{2}}{2r} v^2 u^\mu
+O(\partial^{3})
\nonumber\\
&=&
\frac{1}{T}
\left[
p\,u^\mu
-T^{\mu\nu}u_\nu
-\mu J^\mu
\right]
+S_{\mathrm{HS}}^\mu
+ \frac{\alpha^{2}}{2r} v^2 u^\mu
+O(\partial^{3}) \; ,
	\end{eqnarray}
and the stochastic variables satisfy
\begin{align}
\left\langle
\eta_I^\mu(x)\eta_J^\nu(y)
\right\rangle
={}&
\frac{1}{T} \left[ \left(
\begin{array}{cc}
( \sigma_{\mathrm{DC}}-r\tau ) \Pi^{\mu\nu} + \frac{1}{2} \left(\mathbbm{A}+\mathbbm{A}^\dagger\right)^{\mu\nu}
&
0 \\
0
&
 \left( \frac{\alpha^2}{r\tau}  - \frac{1}{2} D \left(\dfrac{\alpha^2}{r}\right) \right) \Pi^{\mu\nu}
\end{array} \right) \right.
\nonumber\\
& \left. \vphantom{\frac{1}{T}  \left(
\begin{array}{cc}
( \sigma_{\mathrm{DC}}-r\tau ) \Pi^{\mu\nu} + \frac{1}{2} \bigl(\mathbbm{A}+\mathbbm{A}^\dagger\bigr)^{\mu\nu}
&
0 \\
0
&
 \left( \frac{\alpha^2}{r\tau}  - \frac{1}{2} D \left(\dfrac{\alpha^2}{r}\right) \right) \Pi^{\mu\nu}
\end{array} \right)} \qquad
+O(\partial^3) \right]  \delta^{(d+1)}(x-y) \; , \\
\label{eq:noise-covariance}
\mathbbm{A} =& \mathfrak{c}_1u^\mu\nabla^\perp_\nu
+\mathfrak{c}_2\Pi^\mu{}_\nu D
+\mathfrak{c}_3\epsilon^{\mu\rho}{}_{\nu\sigma}
u_\rho B^\sigma \; . 
\end{align}
	\end{subequations}
We have therefore obtained a positive stochastic Maxwell-Cattaneo theory with the second-order holomorphic corrections allowed by the chosen Gaussian embedding.}

\section{Discussion }

{\noindent We have formulated the inverse problem of embedding nonlinear Maxwell-Cattaneo charge transport into a Gaussian Schwinger-Keldysh theory with the modified KMS transformation of \cite{Jain:2023obu} and discovered the relevant constraints one must impose.  We have discussed how one may in principle include higher response terms beyond Gaussian order.}

{\ At the level of the equations of motion, the essential structural choice is to treat the additional vector as intrinsically dissipative: it vanishes, rather than merely becoming time independent, at hydrostaticity. The static generating functional is consequently unchanged, but it no longer exhausts the possible improvements of the entropy current. Those improvements must be specified independently as part of the hydrodynamic data.}

{\ At first derivative order, entropy positivity requires the improvement coefficient $\rho_{v}$ and the conductivity bound \eqref{Eq:PositivityBound}. To subsequently embed the non-linear hydrodynamics in Schwinger-Keldysh, one must impose additional constraints on the transport coefficients as is necessitated by giving the auxiliary field a definitive time reversal parity assignment. With such constraints imposed, the Gaussian fluctuation-dissipation relation then fixes the thermal noise kernel. This establishes the equivalence, at this order, between positivity of the entropy-production quadratic form and positivity of the Gaussian noise kernel.}

{\ At second derivative order, the deterministic constitutive relation does not immediately determine a unique enlarged operator. One can in principle redistribute mixed terms between operator columns without changing their action on the physical thermal-force vector. Hence they leave the constitutive relations and the reduced current response unchanged, while changing the self-adjoint operator and therefore the auxiliary Gaussian representation. This is an ambiguity of the enlarged description, not of the hydrodynamic equations themselves. Imposing the Onsager-Casimir relations removes this ambiguity at second order, but not at third order or beyond. This implies that at higher orders in derivatives, there may be many Gaussian Schwinger-Keldysh embeddings corresponding to the same deterministic dynamics.}

{\  In placing the equations in Langevin canonical form, we have assumed that the additional Maxwell–Cattaneo field admits local and invertible field redefinitions, including redefinitions involving the external sources and their derivatives. This is a natural effective-field-theory assumption if $v^\mu$ is regarded as an auxiliary, non-observable variable, and it underlies our removal of the $\theta_i$ structures in \eqref{Eq:EquationsofMotion}. Nevertheless, in the absence of a microscopic realisation fixing the physical meaning of $v^\mu$, the full class of admissible redefinitions is not determined a priori. A microscopic identification of $v^\mu$, for example with a particular kinetic moment or microscopic operator, could restrict this freedom, in which case some of the operators eliminated in our canonical frame would have to be retained. Our results should therefore be understood as relying on this sensible, but presently unproven, assumption about the field-redefinition freedom of the enlarged effective theory.}

{\ A second, separate issue would arise upon restoring the energy-momentum sector and allowing $T$ and $u^\mu$ themselves to fluctuate. Away from equilibrium, the definition of the hydrodynamic flow is frame dependent, and it has been proposed that correlated redefinitions of the flow and dissipative variables should be regarded as a gauge-like reparametrisation redundancy of fluctuating relativistic hydrodynamics \cite{Dore:2021xqq,Sampaio:2025jtp,Torrieri:2026fjz}. In such a formulation, fluctuations assigned separately to the flow and to relaxation variables need not correspond to independent physical fluctuations. This question does not arise in the fixed-background sector considered here: $T$ and $u^\mu$ are not integrated over, and after eliminating $v^\mu$ the observable theory is expressed entirely in terms of the conserved charge sector. Indeed, \cite{Torrieri:2026fjz} similarly observes that in the pure diffusion limit there is no freedom to redefine the flow frame. Determining how the enlarged Schwinger--Keldysh construction should be modified when fluctuating energy-momentum dynamics is restored is an interesting problem for future work.}

{\ We have not addressed causality and stability of the non-linear stochastic system directly. At the linearised level the constraints $\tau>0$, $r>0$ and $\sigma_{\mathrm{DC}}\geq r\tau$ place both poles of \eqref{Eq:LinearResponsePolesandResidues} in the lower half plane, and the finite relaxation time is precisely the mechanism by which Maxwell-Cattaneo theory renders diffusion causal; a systematic analysis along the lines of \cite{Jain:2023obu,Mullins:2023tjg,Heller:2022ejw,Gavassino:2023myj} within our embedding is left for future work. It would similarly be interesting to compare our intrinsically dissipative vector with the weakly broken symmetry origin of the gapped mode of \cite{Armas:2021vku,Hongo:2024vbj}, and with the one-loop quasi-diffusive correlators of \cite{Abbasi:2025qde}, where fluctuation effects of the relaxational sector become quantitatively accessible.}

{\ The modified affine KMS transformation has also been given a direct stochastic interpretation. For a quadratic action, the Hubbard-Stratonovich representation leads to \eqref{Eq:StochasticEquationsofMotion}, with correlated current and auxiliary noises governed by the same matrix $\mathbbm{N}$. The ratio of a noise history and its transformed history is fixed by the KMS weighting relation derived above, with the thermal derivative of the auxiliary field replaced by the auxiliary field itself. Whether this modified transformation follows from a microscopic notion of local detailed balance, rather than serving only as a consistent effective symmetry, remains an open question.}

\appendix

\section{Contributions from higher response terms}\label{appendix:highernoise}

{\noindent Suppose that the term $\mathcal{K}  \mathcal{O}(X_{a}^3) - \mathcal{O}(X_{a}^3) $ of \eqref{Eq:SKActionTrans} is not KMS closed and produces contributions that enter at orders $X_{a}^{0}$ and $X_{a}^{1}$. As noted previously, this difference cannot contribute to the first term in \eqref{Eq:HSKMSconstraint} as they must necessarily vanish at hydrostaticity. The difference $\mathcal{K}  \mathcal{O}(X_{a}^3) - \mathcal{O}(X_{a}^3) $ can however make contributions to the second and third terms - but these must be equal if we insist that the KMS variation of the action is closed (i.e. $\delta_{\mathrm{KMS}} S=0$) order by order in $X_{a}$. In particular, let us then define
	\begin{eqnarray}
		\label{Eq:HigherTermSKVariation}
		\mathcal{K} \mathcal{O}(X_{a}^3) - \mathcal{O}(X_{a}^3) 
		&=& - i T (\mathcal{L}_{\beta} X)^{T} \mathbbm{F} (\mathcal{L}_{\beta} X) + 2 T X_{a}^{T} \mathbbm{G} (\mathcal{L}_{\beta} X) + \mathcal{O}(X_{a}^2) \; , \qquad
	\end{eqnarray}
where again in the first term we have replaced $X$ by $\Theta X$ in the term $\mathcal{K} \mathcal{O}(X_{a}^3)$ so that
		\begin{eqnarray}
		\label{Eq:Intermediate2}
		&\;& \int d^{3+1} x \;  \left[ - i (\mathcal{L}_{\beta} X)^{T} \left( \begin{array}{c} J \\ V \end{array} \right)_{\mathrm{HS}}  + i ( \mathcal{L}_{\beta} X)^{T} ( \mathbbm{D}_{S} - T ( \mathbbm{N} + \mathbbm{F}) ) ( \mathcal{L}_{\beta} X)  \right. \nonumber \\
		&\;& \left. \hphantom{ + \int d^{3+1} x \;  \left[ \right.} \vphantom{ \left( \begin{array}{c} J \\ V \end{array} \right)_{\mathrm{HS}}} + 2 X_{a} ^{T} \left( - \mathbbm{D}_{S}  + T ( \mathbbm{N} + \mathbbm{G}) \right) (\mathcal{L}_{\beta} X)  + \mathcal{O}(X_{a}^2)  \right] \; . 
	\end{eqnarray}
It follows that the action is KMS closed up to $\mathcal{O}(X_{a}^2)$ if
	\begin{eqnarray}
		\label{Eq:NoiseDissipationRelation}
		\mathbbm{D}_{S} = T ( \mathbbm{N} + \mathbbm{F} ) \; , \qquad \mathbbm{F} = \mathbbm{G} \; .
	\end{eqnarray}
We note that this modified expression has nothing to do with the hydrodynamics being non-linear. Rather, \eqref{Eq:NoiseDissipationRelation} is entirely a property of including non-Gaussian response terms that are not KMS closed. Moreover $\mathbbm{N}$ is still the term entering the thermal Green's function. What \eqref{Eq:NoiseDissipationRelation} tells us is that the relationship between this noise matrix and dissipation is modified in the presence of generic higher response terms (i.e. for generic, explicitly KMS invariant SK actions).}

{\ Let us now review the argument of \cite{Glorioso:2016gsa} relating positivity of the imaginary part of the action to positivity of entropy production. Firstly, if the complete action is KMS invariant then
	\begin{eqnarray}
		\mathscr{L} \stackrel{\mathrm{KMS}}{\longrightarrow} \mathscr{L} - i \nabla_{\mu} \mathscr{V}^{\mu} + (\mathrm{e.o.m}) \; ,
	\end{eqnarray}
where $\mathscr{L}$ is the Lagrangian corresponding to the action \eqref{Eq:SKAction}. Both $\mathscr{L}$ and $\mathscr{V}^{\mu}$ can be expanded order by order in $X_{a}$. The leading terms are given in \eqref{Eq:Intermediate2} and once we impose the modified fluctuation-dissipation \eqref{Eq:NoiseDissipationRelation} we identify
	\begin{eqnarray}
		\left. \partial_{\mu} \mathscr{V}^{\mu} \right|_{X_{a}=0} &=& (\mathcal{L}_{\beta} X)^{T} \left( \begin{array}{c} J \\ V \end{array} \right)_{\mathrm{HS}} + (\mathrm{e.o.m}) = \nabla_{\mu} \left( S^{\mu}_{\mathrm{HS}} + \frac{p u^{\mu}}{T} \right) + (\mathrm{e.o.m}) \; , \qquad
	\end{eqnarray}
where the final identification follows from \eqref{Eq:HydrostaticS}. We can subtract from this the derivative of the canonical entropy current \eqref{Eq:DerivCanonS} to find
	\begin{eqnarray}
		\label{Eq:EntropyConditionfromSK}
		\nabla_{\mu} S^{\mu}  &=& \frac{1}{T}  (\mathcal{L}_{\beta} X)^{T} \mathbbm{D}_{S}  (\mathcal{L}_{\beta} X) + (\mathrm{e.o.m}) \nonumber \\
		&=& (\mathcal{L}_{\beta} X)^{T} \left( \mathbbm{N} + \mathbb{F} \right)  (\mathcal{L}_{\beta} X) + (\mathrm{e.o.m})  \; ,
	\end{eqnarray}
where the second equality follows from \eqref{Eq:NoiseDissipationRelation}. The imaginary part of the action \eqref{Eq:SKAction} must be positive by unitarity
	\begin{eqnarray}
		\mathrm{Im}[S] = T \int d^{3+1}x \; X_{a}^{T} \left( \mathbbm{N} X_{a} + \mathcal{O}'(X_{a}^2) \right) \geq 0
	\end{eqnarray}
where $\mathcal{O}(X_{a}^3) = X_{a}^{T} \mathcal{O}(X_{a}^2)$. In particular, as this must hold for any value of $X_{a}$ it follows that
	\begin{eqnarray}
		\int d^{3+1}x \; X_{a}^{T} \mathbbm{N} X_{a}  \geq 0 \; . 
	\end{eqnarray}
All that remains is to show that $\mathbbm{N} \geq 0$ implies that the right hand side of \eqref{Eq:EntropyConditionfromSK} is positive. To do this, following \cite{Glorioso:2016gsa}, we note that the derivative order of $\mathbbm{F}$ is at least $\mathcal{O}(\partial)$ (see the definition in \eqref{Eq:HigherTermSKVariation}) while the lowest derivative order of $\mathbbm{N}$ is $\mathcal{O}(\partial^{0})$. We can therefore perturbatively redefine the vectors in \eqref{Eq:EntropyConditionfromSK} to ensure that the raw $\mathbbm{N}$ matrix is sandwiched between two modified vectors. For example 
	\begin{subequations}
	\begin{eqnarray}
		\nabla_{\mu} S^{\mu}
			&=&  (\mathcal{L}_{\beta} X)^{T} \left( \mathbbm{N}_{(0)} + \mathbbm{N}_{(1)} + \ldots + \mathbbm{F}_{(1)} + \mathbbm{F}_{(2)} + \ldots \right)  (\mathcal{L}_{\beta} X) \nonumber \\
			&\;& + (\mathrm{e.o.m}) \\
			&=&  Y^{T} \left( \mathbbm{N}_{(0)} + \mathbbm{N}_{(1)} + \ldots + \mathbbm{N}_{(m)} \right) Y  + (\mathrm{e.o.m})  + \mathcal{O}(\partial^{m+1}) \; , \qquad \\
					Y &=& (\mathcal{L}_{\beta} X) + \frac{1}{2} \mathbbm{N}_{(0)}^{-1} \mathbbm{F}_{(1)} (\mathcal{L}_{\beta} X)  + \ldots  \; , 
	\end{eqnarray}
	\end{subequations}
where $_{(i)}$ indicates terms at derivative order $i$ and we have required the additional assumption of $\mathbbm{N}_{(0)}$ being invertible. One proceeds order by order until $\mathbbm{F}$ is completely eliminated to the relevant number of derivatives. In this way we see that positivity and invertibility of $\mathbbm{N}$ can be used to imply that the entropy current is positive definite on-shell i.e.
	\begin{eqnarray}
		\nabla_{\mu} S^{\mu}  &=&  (\mathcal{L}_{\beta} X)^{T} \left( \mathbbm{N} + \mathbb{F} \right)  (\mathcal{L}_{\beta} X) \nonumber \\
		&=&  Y^{T} \left( \mathbbm{N}_{(0)} + \mathbbm{N}_{(1)} + \ldots + \mathbbm{N}_{(m)} \right) Y \geq 0 \; ,
	\end{eqnarray}
which completes the demonstration.}

\section{$PT$ table}\label{appendix:PTtable}

\begin{table}[!h]
    \centering
    \renewcommand{\arraystretch}{1.25}
    \begin{tabular}{c c c c}
        \hline\hline
        Quantity & $P$ & $T$ & $PT$ \\
        \hline
        $t$                               & $+$ & $-$ & $-$ \\
        $x^i$                             & $-$ & $+$ & $-$ \\
        $\mu$                             & $+$ & $+$ & $+$ \\
        $J^0$                             & $+$ & $+$ & $+$ \\
        $J^i$                             & $-$ & $-$ & $+$ \\
        $v^i$                             & $-$ & $-$ & $+$ \\
        $V^i$                             & $-$ & $-$ & $+$ \\
        $E^i$                             & $-$ & $+$ & $-$ \\
        $\nabla_\perp^i\mu$               & $-$ & $+$ & $-$ \\
        $ E^i-\nabla_\perp^i\mu$ & $-$ & $+$ & $-$ \\
        $\mathcal B^i$                    & $+$ & $-$ & $-$ \\
        $\partial_t$                      & $+$ & $-$ & $-$ \\
        $\partial_i$                      & $-$ & $+$ & $-$ \\
        $D\equiv u^\mu\nabla_\mu$         & --- & --- & $-$ \\
        $\nabla_\perp^i$                  & --- & --- & $-$ \\
        $\epsilon^{\mu\rho}{}_{\nu\sigma}
          u_\rho\mathcal B^\sigma$        & --- & --- & $-$ \\
        \hline\hline
    \end{tabular}
    \caption{
        Discrete-symmetry assignments in the fixed-background rest-frame
        sector.  The spacetime map is
        $\vartheta(t,\mathbf{x})=(-t,-\mathbf{x})$.
        The entries denote the transformation of the complete object,
        including its tensor indices and its argument. 
    }
    \label{tab:PT-assignments}
\end{table}

{\noindent We take the spacetime involution appearing in the dynamical KMS
transformation to be the full \(PT\) inversion,
\begin{equation}
    \vartheta x=(-t,-\mathbf{x}) .
\end{equation}
For any field \(X\), we write
\begin{equation}
    (\Theta X)(x)
    =\eta_X\,X(\vartheta x),
\end{equation}
where \(\eta_X\) is the intrinsic \(PT\) eigenvalue after accounting for
the tensor indices of \(X\).  In particular, the two components of the
enlarged dissipative sector have equal \(PT\) parity,
\begin{equation}
    \eta_J=\eta_v=+1 .
\end{equation}
Parity is included in \(\vartheta\) and is therefore not applied a
second time through \(\Theta\).  Differential operators and background
fields are transformed as complete operators; for example,
\begin{equation}
    D^{PT}=-D,
    \qquad
    (\nabla_\perp^\mu)^{PT}=-\nabla_\perp^\mu,
    \qquad
    (E^\mu-\nabla_\perp^\mu\mu)^{PT}
    =-(E^\mu-\nabla_\perp^\mu\mu).
\end{equation}
}

\begin{table}[!h]
    \centering
    \renewcommand{\arraystretch}{1.25}
    \begin{tabular}{c c}
        \hline\hline
        Schwinger--Keldysh field & Intrinsic $PT$ eigenvalue \\
        \hline
        $B_{r\,0}$       & $+1$ \\
        $B_{r\,i}$       & $+1$ \\
        $B_{a\,0}$       & $+1$ \\
        $B_{a\,i}$       & $+1$ \\
        $\zeta_r^i$      & $+1$ \\
        $\zeta_a^i$      & $-1$ \\
        \hline\hline
    \end{tabular}
    \caption{
        Intrinsic \(PT\) assignments used in the enlarged
        Schwinger--Keldysh field space. The component transformations
        induced by spacetime inversion are already included through
        \(\vartheta x\).
    }
    \label{tab:SK-PT-assignments}
\end{table}

\bibliographystyle{JHEP}
\bibliography{references}
\end{document}